\documentclass[%
reprint,
superscriptaddress,
amsmath,amssymb,
aps,
prl,
]{revtex4-1}

\usepackage{float}
\usepackage{graphicx}
\usepackage{dcolumn}
\usepackage{bm}
\usepackage{hyperref}
\usepackage{xcolor}
\usepackage[T1]{fontenc}
\hypersetup{
  colorlinks,
  citecolor=blue,
  linkcolor=blue,
  urlcolor=blue}

\usepackage{braket}
\usepackage{upgreek}

\newcommand{\Ge}[1]{\texorpdfstring{\textsuperscript{#1}Ge}{{#1}Ge}}

\def\markup{0}
\ifnum\markup=1
\newcommand{\red}[1]{\textcolor{red}{#1}}
\else
\newcommand{\red}[1]{#1}
\fi

\begin{document}

\preprint{APS/123-QED}

\title{Hyperfine spectroscopy and fast, all-optical arbitrary state initialization and readout of a single, ten-level \Ge{73} vacancy nuclear spin qudit \red{in diamond}}

\author{C. Adambukulam}
\email[Corresponding author: ]{c.adambukulam@unsw.edu.au}
\affiliation{
School of Electrical Engineering and Telecommunications, University of New South Wales, Kensington, NSW 2052, Australia
}

\author{B. C. Johnson}
\affiliation{
School of Science, RMIT University, Melbourne, VIC 3001, Australia
}

\author{A. Morello}
\affiliation{
School of Electrical Engineering and Telecommunications, University of New South Wales, Kensington, NSW 2052, Australia
}
 
\author{A. Laucht}
\email[]{a.laucht@unsw.edu.au \newline}
\affiliation{
School of Electrical Engineering and Telecommunications, University of New South Wales, Kensington, NSW 2052, Australia
}

\date{\today}

\begin{abstract}
A high-spin nucleus coupled to a color center can act as a long-lived
memory qudit in a spin-photon interface. The germanium vacancy (GeV) in diamond has attracted recent attention due to its excellent spectral properties and provides access to the 10-dimensional Hilbert space of the $I = 9/2$ \Ge{73} nucleus. Here, we observe the \Ge{73}V hyperfine structure, perform nuclear spin readout, and optically initialize the \Ge{73} spin into any eigenstate on a $\mu$s-timescale and with a fidelity of up to \red{$\sim 84\%$}. Our results establish \Ge{73}V as an optically addressable high-spin quantum platform for a high-efficiency spin-photon interface as well as for foundational quantum physics and metrology.
\end{abstract}

\maketitle


Color centers are attractive candidates for light-matter qubit interfaces, as required to realize a quantum network~\cite{Ruf2021, Togan2010, Nguyen2019, Hensen2015}. In such systems, electrons bound to the color center potential provide a localized spin (or, matter) qubit, while spin selective optical transitions enable spin initialization~\cite{Rogers2014}, readout~\cite{Chen2022}, and spin-photon entanglement~\cite{Nguyen2019}. Group-IV split vacancy defects in diamond are particularly \red{promising} owing to their excellent spin~\cite{Sukachev2017, Debroux2021} and spectral~\cite{Chen2022, Rogers2014B} properties, even when incorporated into nanostructures~\cite{Trusheim2020}. \red{The group-IV defect of interest in this Letter, the germanium vacancy (GeV), has demonstrated an optical life- and coherence time of $\tau_{\rm r} \sim 5.9$~ns and $T_{2, \rm opt}^* \sim 9.5$~ns~\cite{Chen2022} in addition to electron spin coherence times exceeding $\sim 20$~ms~\cite{Senkalla2023}.}

We can expand the quantum applications of \red{color centers} by including hyperfine coupled nuclear spins with coherence times that greatly exceed that of the electron \cite{Waldherr2012}. For diamond-based color centers, work has typically focused on \red{weakly coupled \textsuperscript{13}C spins}~\cite{Dutt2007, Bradley2019} with several fundamental demonstrations regarding quantum networks having been performed with \red{them}~\cite{Kalb2017, Tsurumoto2019, Pompili2021, Hermans2022}. However, the non-deterministic inclusion of the \textsuperscript{13}C isotope \red{presents additional challenges with regards to locating \textsuperscript{13}C nuclear spins. In contrast, each of the group-IV elements has at least one isotope with a nuclear spin. High cooperativity between a group-IV defect and a nanophotonic cavity allows for these intrinsic nuclear spins to serve as quantum memories, thereby bypassing the need to locate \textsuperscript{13}C spins. This has been demonstrated for the \textsuperscript{29}SiV~\cite{Stas2022}.}

\begin{figure}[t]
\includegraphics{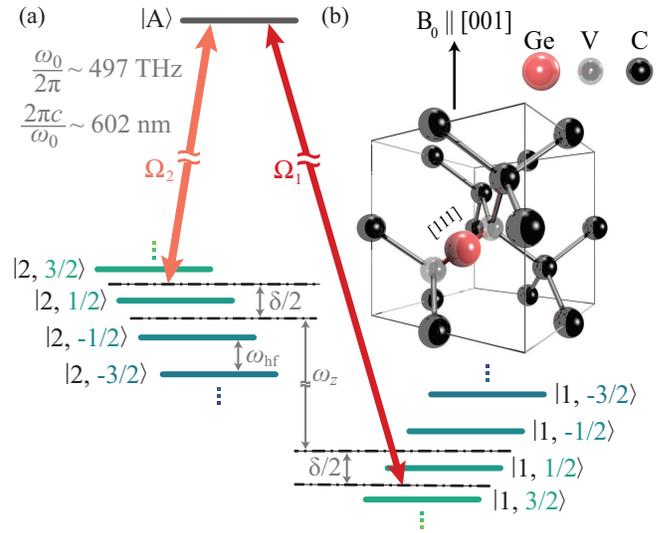}
\caption{\label{fig:1} 
(a) Level structure of the \Ge{73}V, with CPT driving lasers denoted by the orange and red arrows. The optical Rabi frequency between $\ket{\rm A}$ and $\ket{1}$ and between $\ket{\rm A}$ and $\ket{2}$ is $\Omega_1$ and $\Omega_2$. 
\red{The lasers have frequencies approximately equal to the GeV zero-phonon line of $\omega_0/2\pi \sim 497$~THz or $\sim 602$~nm. Additionally, $c$, $\delta$, $\omega_{\rm z}$ and $\omega_{\rm hf}$ refer to the the vacuum speed of light, two laser detuning, Zeeman splitting and hyperfine splitting.} (b) Diagram of the GeV with the orientation of the magnetic field, $B_0$ shown. The angle between $B_0$ and $[111]$ is $\sim 54.7^\circ$. Ge, V, and C refer to germanium, vacancy, and carbon, respectively.}
\end{figure} 

Of all the group-IV isotopes, including \textsuperscript{13}C, only \Ge{73} has a high nuclear spin ($I > 1/2$). In fact, the $I = 9/2$ \Ge{73} spin spans a 10-dimensional Hilbert space. Several proposed schemes~\cite{Gross2021, Petiziol2021, Gross2021B} map a two-level qubit onto a nuclear spin qudit and leverage the redundancy of a large Hilbert space for error correction. Such schemes could significantly improve the storage time of a nuclear spin-based quantum memory. In addition, high-spin nuclei are a potential platform for investigating several fundamental questions in quantum mechanics; from quantum chaos~\cite{Mourik2018} to the reality of the wavefunction~\cite{Barrett2014}, and as well as for quantum metrology, where spin analogues to non-classical states~\cite{Korkmaz2016} of light may improve sensor gain. However, the experimental study of solid-state high spin-nuclei has been limited, with few experimental platforms available~\cite{Asaad2020, Chekhovich2020, Thiele2014, De_Fuentes2023}. This extends further when considering optically active defects as other centers with access to a high spin nucleus ~\cite{Hendriks2022, Bosma2018, Yang2022Zeeman, Viitaniemi2022} have either yet to demonstrate single defect optical addressability or do not exhibit comparable spin and spectral properties.

In this Letter, we experimentally establish the negatively charged \Ge{73}V in diamond as a powerful, optically addressable 10-level nuclear spin platform in the solid state. We perform coherent population trapping (CPT) experiments to resolve the hyperfine structure of the \Ge{73}V. Using pulsed CPT, we readout the nuclear spin state and observe rapid nuclear spin diffusion that arises from nuclear spin non-conserving optical relaxation. We leverage these relaxation processes to perform fast, high-fidelity all-optical initialization of the \Ge{73} spin into any of its eigenstates.


\begin{figure}[t]
\includegraphics{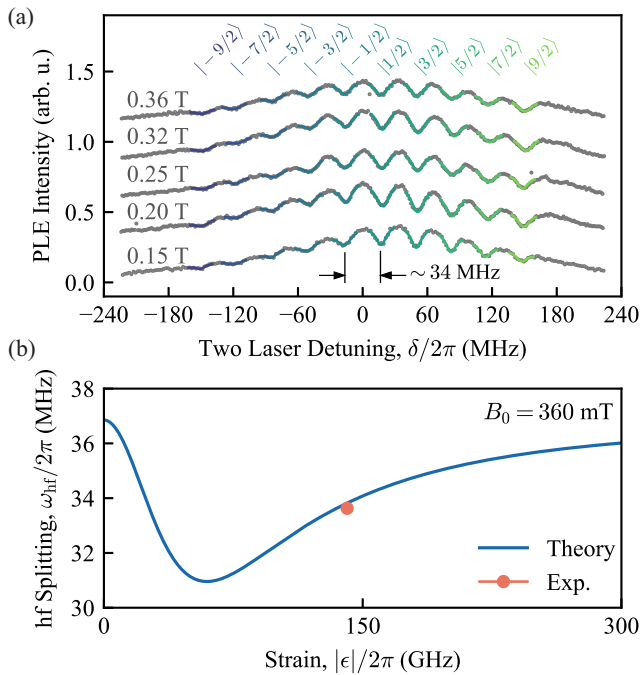}
\caption{\label{fig:2} 
(a) Coherent population trapping spectra with $B_0$ varied (grey text). \red{The CPT power ranges from $\sim 220$~nW to $\sim 420$~nW or equivalently, between 3 and 6 times the saturation power of A1 -- which was measured to be $p_{\rm sat} = 70 \pm 5$~nW}. The data (grey dots) is fit (colored lines) to ten \red{copies of a Lorentzian, $A/(1+(\delta-\delta_0)/\gamma)^2) + C(\delta - \delta_0) + D$~\cite{SuppMat}. Each copy is centered on a particular \textit{dip}} that corresponds to one of the ten nuclear spin eigenstates ordered $\ket{-9/2}$ to $\ket{9/2}$ from left to right. 
(b) Calculated strain dependence of the hyperfine splitting when $B_0 \parallel [001]$ and $A/2\pi = 36.98$~MHz. The orange point indicates the experimentally measured hyperfine splitting.}
\end{figure}

Fig.~\ref{fig:1}(a) shows the level structure of the \Ge{73}V where $\ket{1}$ and $\ket{2}$ refer to the electron spin-down and -up eigenstates of the ground state (GS) and $\ket{\rm A}$ to the electron spin-down eigenstate of the optically excited state (ES). In an off-axis magnetic field, $B_0 \nparallel [111]$ [see Fig.~\ref{fig:1}(b)] the transition between $\ket{2}$ and $\ket{\rm A}$ is weakly allowed resulting in a $\Lambda$-system. We perform CPT~\cite{Rogers2014, pingault2014all, Agapev1993, Xu2008} by applying two equal power \red{(as imposed by the experimental setup~\cite{SuppMat})} lasers to simultaneously drive the $\ket{1} \leftrightarrow \ket{\rm A}$ and $\ket{2} \leftrightarrow \ket{\rm A}$ (denoted A1 and A2) transitions with Rabi frequencies $\Omega_1$ and $\Omega_2$ as shown in Fig.~\ref{fig:1}(a).  \red{In a system where no hyperfine coupled nuclear spin is present and when the two laser detuning, $\delta/2\pi = 0$~MHz [defined in Fig.~\ref{fig:1}(a) and Ref.~\onlinecite{SuppMat}], the system steady state is the so-called \textit{dark-state}; a superposition of $\ket{1}$ and $\ket{2}$ which produces a \textit{dip} in a photo-luminescence excitation (PLE) spectrum~\cite{arimondo1996v}. A hyperfine coupled nuclear spin introduces a nuclear state dependent shift $\sim m \omega_{\rm hf}$ - for a nuclear state $\ket{m}$ - to the CPT resonance such that it may no longer be at $\delta/2\pi = 0$~MHz}. The Hamiltonian that describes this interaction is,
\begin{equation}
    H_{\rm hf} = \red{\frac{A_{\perp}}{2} (S_+ I_- + S_- I_+)} + A_{\parallel} S_z I_z,
\end{equation} where $S_i$ ($I_i$) for $i \in \{x, y, z\}$ are the electron (nuclear) spin operators, $S_\pm = S_x \pm iS_y$ (and likewise, $I_\pm = I_x \pm iI_y$) and $A_{\parallel}$ ($A_{\perp}$) is the longitudinal (transverse) hyperfine coupling. As the GS wavefunction parity is even, the isotropic Fermi contact interaction dominates~\cite{Rogers2014, Karim2023} and $A = A_{\parallel} \sim A_{\perp}$. 

In Fig.~\ref{fig:2}(a), we plot the results of a CPT measurement on a single \Ge{73}V center. \red{Note, the single laser detuning and sample temperature is $\sim 0$~MHz and $\sim 25$~mK in this and all experiments described hereafter~\cite{SuppMat}. Furthermore, as shown in Fig.~\ref{fig:1}(a), we perform CPT by sweeping $\delta$ and changing frequency of both lasers.}  We observe ten $B_0$ independent dips, split by $\omega_{\rm hf}/2\pi = 33.81 \pm 0.05$~MHz, that correspond to the ten \Ge{73} spin eigenstates. As with the \textsuperscript{29}SiV~\cite{Pingault2017}, $\omega_{\rm hf}$ depends on the strain and $B_0$ orientation [see Fig.~\ref{fig:2}(b)]. By fitting the measured $\omega_{\rm hf}$ to the GeV Hamiltonian, we extract $A_{\parallel}/2\pi = 36.98 \pm 0.06$~MHz assuming an orbital Zeeman effect of $\sim 0.1 \gamma_e B_0/2$~GHz~\cite{Hepp2014B}. The measured value is in good agreement with its \textit{ab initio} prediction in Ref.~\onlinecite{Karim2023}. The strain dependence of $\omega_{\rm hf}$ arises from strain induced mixing of the electron spin and orbital degrees of freedom (see Ref.~\onlinecite{SuppMat}) that occurs when strain, $|\epsilon|$ is comparable to the spin-orbit coupling, $\lambda \sim 165$~GHz~\cite{Maity2018}. \red{At low excitation powers, the CPT \textit{dip} width is given by $\propto T_2^{*-1}$. In practice, we choose excitation powers such that, $\Omega_1, \Omega_2 \gg T_2^{*-1}$ to ensure \textit{dip} visibility~\cite{Agapev1993}. This lowers the CPT resolution by way of power broadening (from several $100$~kHz to several MHz) and thus,} we cannot measure the $\sim$~kHz~\cite{Karim2023} magnitude shifts produced by the second order hyperfine and quadrupole interactions.

We apply perturbation theory to the \Ge{73}V hyperfine structure (see Ref.~\onlinecite{SuppMat}) to understand its strain dependence. We find that the impact of strain on the effective transverse hyperfine coupling is analogous to its impact on the electron g-factor~\cite{SuppMat, Hepp2014, Hepp2014B}. More explicitly, and up to the second order in $\epsilon$, $A_\perp^{\rm eff} = 4|\epsilon|A_\perp/\lambda$. Given that the nuclear Zeeman effect is weak, the hyperfine interaction entirely dominates the dynamics of the nuclear spin.


\begin{figure*}[t]
\centering
\includegraphics[width=\textwidth]{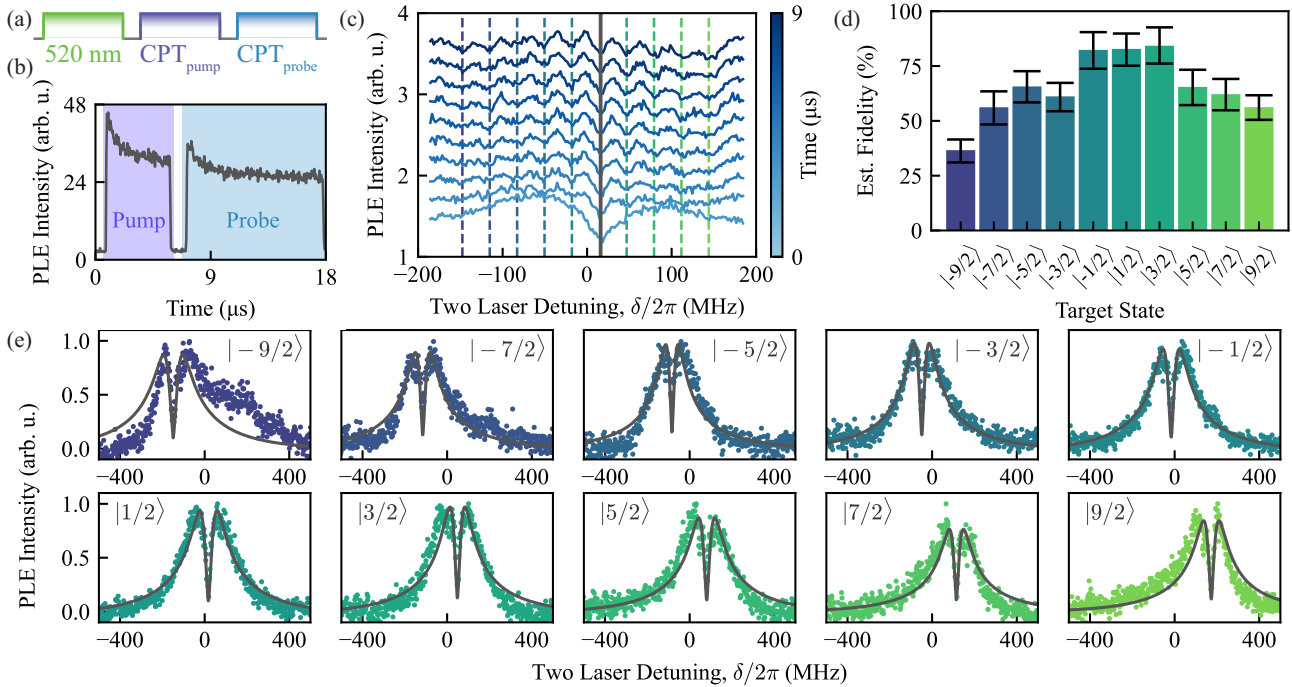}
\caption{\label{fig:3} 
(a) Pulse sequence for a nuclear pump-probe experiment. \red{CPT\textsubscript{pump} and CPT\textsubscript{probe} refer to two, CPT pulses - each pulse consisting of two simultaneously applied lasers - where $\delta$ is either fixed and resonant to a CPT \textit{dip} (pump) or varied during the experiment (probe). Note, \textit{pump} and \textit{probe} exclusively refer to the effect of these pulses on the nuclear spin.}
(b) A histogram of the photon arrival times measured during a nuclear pump-probe experiment. \red{CPT\textsubscript{pump} is resonant to $\ket{1/2}$ while CPT\textsubscript{probe} is resonant to $\ket{5/2}$.}
(c) Time resolved PLE intensity during a probe pulse. The CPT\textsubscript{pump} pulse is resonant to $\ket{1/2}$ as denoted by the solid grey line. The dashed, colored lines indicate the resonances of the other nuclear spin eigenstates. 
(d) The \red{estimated} initialization fidelity achieved for each nuclear spin eigenstate. \red{Here, the error bars refer to the $95\%$ confidence interval.}
(e) CPT spectra measured by the nuclear pump-probe technique with the PLE intensity measured during the first \red{$250$}~ns of the CPT\textsubscript{probe} pulse. The data (coloured dots) has been fit to the steady state of a CPT Lindbladian (grey lines, see Ref.~\onlinecite{SuppMat}). The CPT\textsubscript{pump} frequency is indicated by the color of the dots which corresponds to the color of dashed lines in (b) with the associated nuclear spin state labelled adjacent to the data (grey text).}
\end{figure*}

\begin{figure}[t]
\centering
\includegraphics{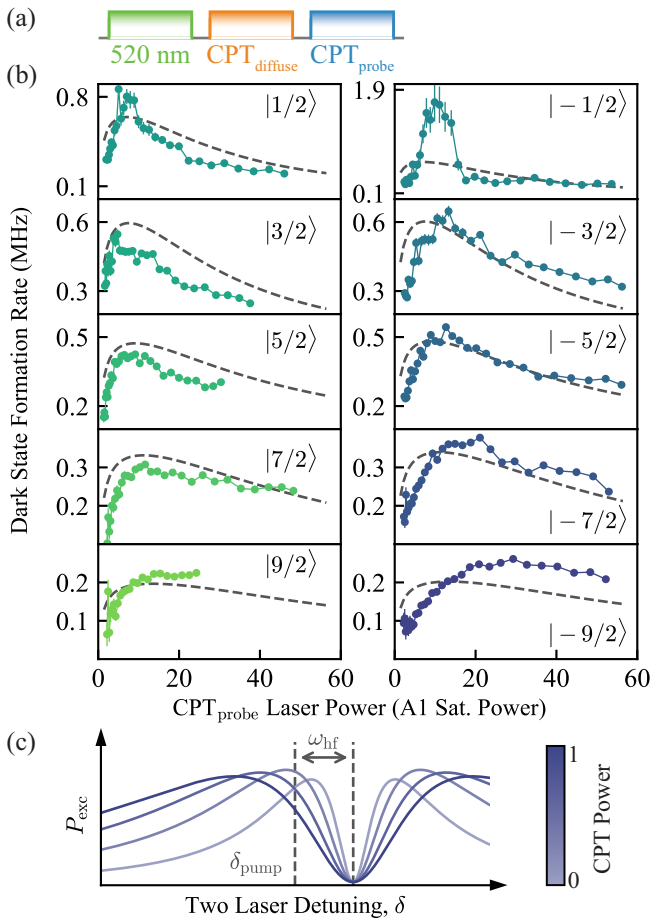}
\caption{\label{fig:4} Nuclear spin initialization time. 
(a) The pulse sequence used to measure the nuclear spin initialization time. Here, CPT\textsubscript{diffuse} refers to a completely off-resonant, $\delta = 0$~MHz CPT pulse. 
(b) \red{The power dependence of the nuclear spin initialization rates for each nuclear spin eigenstate. Here, the laser power is expressed in terms of the A1 saturation power, $p_{\rm sat} = 70 \pm 5$~nW.} The colored points denote the measured data while the grey, dashed lines plot the theoretical initialization rate extracted from a rate model.
(c) \red{Schematic diagram of the excitation probability, $P_{\rm exc}$, at the beginning of a CPT\textsubscript{pump}/CPT\textsubscript{probe} pulse. Here, $\delta$ is set to pump the nuclear spin eigenstate adjacent to the initial state of the system resulting in a CPT detuning of $\omega_{\rm hf}$ from the resonance. Note, when the CPT power is large, increasing it further reduces the excitation rate  as the \textit{dip} broadens and consequently suppresses nuclear spin diffusion.}}
\end{figure}

We perform a pump-probe experiment with the pulse sequence shown in Fig.~\ref{fig:3}(a). We apply a CPT-pump pulse where $\delta$ is resonant with a particular nuclear state, followed by a CPT-probe pulse where $\delta$ is varied for nuclear spin readout~\cite{Golter2013}. \red{The $\sim 520$~nm pulse stabilizes the GeV charge state~\cite{Chen2019}.} The time-resolved PLE, collected during the probe pulse, is shown in the histograms in Fig.~\ref{fig:3}(b) and (c). In Fig.~\ref{fig:3}(c), this is done with the pump-pulse resonant with the nuclear spin eigenstate, $\ket{1/2}$. The PLE measured \red{at the start of} the probe pulse, then verifies that the nuclear spin was pumped into $\ket{1/2}$. This is extended to the entire nuclear spin Hilbert space as shown in Fig.~\ref{fig:3}(e) where the different panels show initialization into all ten nuclear spin eigenstates. The \red{estimated} initialization fidelities for each eigenstate are plotted in Fig.~\ref{fig:3}(d) and are computed from the data in Fig.~\ref{fig:3}(e). We \red{estimate typical fidelities of $\sim 65\%$ with the highest fidelity of $84^{+9}_{-8}\%$ deduced for $\ket{3/2}$}. 

The initialization mechanism is as follows; at the start of a target-state-resonant CPT pulse, the nuclear spin is in an arbitrary state and thus, the CPT pulse is not resonant with the system. Consequently, optical excitation occurs and the subsequent relaxation flips the nuclear spin. The nuclear spin continues to flip, or diffuse, until it reaches the target state upon which formation of the CPT dark-state terminates the process. This protocol has found applications in Overhauser field cooling of self-assembled quantum dots~\cite{Ethier-Majcher2017} and NV centers~\cite{Togan2011}. In those systems, nuclear spin diffusion arises from the transverse hyperfine interaction, \red{$A_\perp(S_+I_- + S_-I_+)$}, which generates electron-nuclear spin flip-flops. For highly strained group-IV defects,  $\omega_z$ suppresses this term~\cite{Issler2010} resulting in nuclear spin raising and lowering rates of $A_\perp^2 \Gamma_r/4\omega_z^2$ and $A_\perp^2 \Gamma_r/4\eta \omega_z^2 \sim 10-100$~Hz. Here, $\Gamma_r$ is the optical relaxation rate and $\eta \approx \Omega_1^2/\Omega_2^2$ is the branching ratio. In Fig.~\ref{fig:4}(c), we plot the relationship between dark-state formation rate and laser power during the nuclear spin initialization protocol. To measure this, we modify the nuclear pump-probe pulse sequence [see Fig.~\ref{fig:3}(a)] to a diffuse-probe sequence [see Fig.~\ref{fig:4}(b)] wherein the two laser detuning of the first pulse is $\delta/2\pi = 0$~MHz and completely off-resonant to any nuclear spin eigenstate. \red{This pumps the nuclear spin into a mixed state from which we may initialize a specific nuclear spin projection. By fixing $\delta$ during the CPT-diffuse pulse, the state populations of the resulting mixed state are the same between the various experimental runs.} We measure initialization rates exceeding $100$~kHz for all ten eigenstates; far beyond what the transverse hyperfine could produce.

We extend our perturbative model of the \Ge{73}V hyperfine structure to its dipole operators~\cite{SuppMat}. We find that the difference in GS and ES strain modulates the nuclear spin flipping transition strength. This is analogous to the electron spin flipping transition, A2; its intensity being strain modulated. The largest contribution to nuclear spin diffusion is the difference in the GS and ES nuclear spin quantization axes. As the hyperfine interaction dominates, the nuclear spin quantization direction depends on that of the electron and on the components of the hyperfine tensor. Whereas the GS hyperfine tensor is dominated by the isotropic Fermi contact interaction, the ES hyperfine tensor is expected to be dominated by the anisotropic dipolar term~\cite{Harris2023}. This, in combination with differing GS and ES g-tensors results in different GS and ES nuclear spin quantization axes thereby increasing the likelihood of relaxation induced nuclear spin flips. 

We construct a rate model of the nuclear spin pumping process~\cite{SuppMat}, to gain insight into its excitation power dependence [see Fig. \ref{fig:4}~(a)]. In particular, we note that rather than saturating, the pumping rate decreases after some optimum power. This is a consequence of power broadening \red{[see Fig.~\ref{fig:4}(c)]}. Wider CPT dips suppress optical excitation and slow the nuclear spin diffusion process. Deviation of the rate model from the experiment is due to sensitivity of the modelled nuclear spin quantization axis to the unknown value of the ES hyperfine and in addition to, errors and fluctuations in $\delta$ for both CPT pulses in the several hours of averaging required to perform the experiment.

Given the nuclear spin diffusion during the probe pulse \red{[see Fig.~\ref{fig:3}(c), where the remaining nine \textit{dips} appear after the CPT-probe pulse is applied for several $\mu$s]}, the fidelity estimates shown in Fig.~\ref{fig:3}(d) are conservative. \red{Additionally, PLE intensity drift during the measurement further limits the accuracy of fidelity estimation}. The low initialization fidelity into $\ket{-9/2}$ is likely due to the difficulty in ascertaining the corresponding $\delta$ given its low CPT contrast \red{and in addition to fast nuclear spin diffusion away from $\ket{-9/2}$ during the CPT-probe pulse}. Fidelity is limited by errors in $\delta$ - an effect amplified by power broadening. The fundamental limit to fidelity is the probability of re-excitation from the dark-state, given by the ratio of the optical lifetime and the electron $T_2^*$~\cite{SuppMat}. The initialization fidelity stands to be improved through the optimization of $T_2^*$ by isotopic purification of the diamond and by aligning $B_0$ to $[111]$~\cite{Sukachev2017, Senkalla2023} while minimizing errors in $\delta$. 

CPT-based nuclear spin pumping~\cite{Jamonneau2022} provides access to nuclear spin systems where the difference between the GS and ES hyperfine splittings does not exceed the optical linewidth~\cite{Parker2023} and presents less overhead than traditional initialization schemes such as electron initialization followed by a SWAP gate~\cite{Stas2022}, dynamic nuclear polarization~\cite{Poggiali2017}, or measurement-based initialization \cite{Robledo2011}. Moreover, the ability to polarize into any nuclear eigenstate allows one to bypass radio-frequency (RF) control of the nuclear spin during initialization. This is highly desirable given the large, ten-dimensional nuclear spin Hilbert space and the low \Ge{73} gyromagnetic ratio, $\gamma_n \sim 1.5$~MHz/T. For example, assuming a high RF power to magnetic field conversion efficiency as achieved in Ref.~\onlinecite{Vallabhapurapu2021Fast} and a dilution refrigerator power budget of $100$~$\rm{\mu}$W, we would expect nuclear Rabi frequencies on the order of $\sim 100$~Hz and significantly lower initialization rates. In contrast, we show $\mu$s-timescale initialization rates comparable to that of the electron spin (see Ref.~\onlinecite{SuppMat}). On the other hand, the fast nuclear spin diffusion makes nuclear spin readout with CPT difficult, as the number of collected photons per shot is $\ll 1$. 
\red{However, we expect that the nuclear spin diffusion rate may be lowered by aligning $B_0$ with the defect axis and in doing so, improve CPT-based nuclear spin readout. Additionally,} alternative readout mechanisms exist. Namely, a nuclear spin conditional operation on the electron and its subsequent readout \cite{Stas2022, Neuman2010} which may also enable single-shot nuclear spin readout.


The long coherence times of solid-state nuclear spins \cite{Waldherr2012} make them an invaluable resource for color-center-based spin-photon interfaces, which has motivated recent investigations into the intrinsic nuclear spins of group-IV defects~\cite{Stas2022, Parker2023, Harris2023}. In this article, we observe the hyperfine structure of the \Ge{73}V and measure $A_\parallel \sim 37$~MHz. Furthermore, we demonstrate optical readout and initialization of the \Ge{73} spin, quickly and without requiring microwave or RF magnetic fields. This constitutes a feature that significantly improves the feasibility of addressing the nuclear spin, given its large Hilbert space and low gyromagnetic ratio. Our work will fundamentally enable a near-term demonstration of coherent control of the \Ge{73} spin, either via all-optical~\cite{Vallabhapurapu2022} or magnetic methods. In the long term, the \Ge{73}V system could be deployed as an optically-accessible qudit for quantum information processing~\cite{Wang2020}, as a platform to explore quantum chaos~\cite{Mourik2018} or to generate non-classical spin states of metrological interest~\cite{Korkmaz2016}.
\newline

\begin{acknowledgments}
We would like to thank Blake Regan for his assistance with sample preparation and Hyma H. Vallabhapurapu for useful discussions. We acknowledge funding from the Australian Research Council (Grant no. CE170100012). This work used the facilities of the Australian National Fabrication Facility (ANFF). We acknowledge access and support to NCRIS facilities (ANFF and the Heavy Ion Accelerator Capability) at the Australian National University. C. A. and A. L. acknowledge support from the University of New South Wales Scientia program.
\end{acknowledgments}

\bibliography{references}

\end{document}


\preprint{APS/123-QED}

\title{Supplemental Material for: \texorpdfstring{\\}{} Hyperfine spectroscopy and fast, all-optical arbitrary state initialization and readout of a single, ten-level \texorpdfstring{$^{73}$}{73-}Ge vacancy nuclear spin qudit in diamond}

\author{C. Adambukulam}
\affiliation{%
 School of Electrical Engineering and Telecommunications, University of New South Wales, Kensington NSW, Australia 2052
}%

\author{B. C. Johnson}
\affiliation{%
School of Science, RMIT University, Melbourne VIC, Australia 3001
}%

\author{A. Morello}
\affiliation{%
 School of Electrical Engineering and Telecommunications, University of New South Wales, Kensington NSW, Australia 2052
}%
 
\author{A. Laucht}
\affiliation{%
 School of Electrical Engineering and Telecommunications, University of New South Wales, Kensington NSW, Australia 2052
}%

\maketitle

\tableofcontents

\section{Theoretical Description of the Germanium Vacancy}

\subsection{Electronic Structure}

Here, we briefly summarise the description of group-IV split vacancy defects developed in Ref.~\cite{Hepp2014}. In the $-1$ charge state, the GeV ground and excited states consist of two orbitals, the degeneracy of which, is lifted by spin-orbit coupling, crystal strain, the dynamic Jahn-Teller effect (DJT) and the Zeeman effect. If we choose our basis to be the eigenstates of the $z$-component of the orbital angular momentum operator $\{\ket{e_{+}}, \ket{e_{-}}\}$, we may write the orbital angular momentum operators as,
\begin{equation}
    L_x = 0, \,\, L_y = 0, \,\, \textrm{and} \,\, L_z = \begin{pmatrix}
        1 & 0 \\
        0 & -1 
    \end{pmatrix}.
\end{equation} Note, the form the Hamiltonian takes in the ground and the excited state is indistinguishable and thus, we do not explicitly differentiate between the two. However, when that distinction is necessary, we denote the ground state by a subscript $g$ and the excited state by a subscript $u$. Additionally, we define our coordinate system such that $z$ refers to the high symmetry axis of the $D_{3d}$ point group and is oriented along $[111]$. The orbital Zeeman effect is described by the term,
\begin{equation}
    q \gamma_o B_z L_z.
\end{equation} Here, $\gamma_o \sim 14$~GHz/T is orbital gyromagnetic ratio, $B_z$ is the magnetic field applied along $z$ and $q$ is the orbital Zeeman quenching parameter and is $\approx 0.1$. By symmetry, the effect of crystal strain can be reduced to three operators: 
\begin{equation}
    \mathcal{A} = \kappa \begin{pmatrix}
                            1 & 0 \\
                            0 & 1 
                        \end{pmatrix}, \,\, 
    \mathcal{E}_{x} = \alpha \begin{pmatrix}
                            0 & 1 \\
                            1 & 0 
                        \end{pmatrix} \,\, \textrm{and}	\,\,
    \mathcal{E}_{y} = \beta \begin{pmatrix}
                            0 & -i \\
                            i & 0 
                        \end{pmatrix},
\end{equation} where $\kappa$, $\alpha$ and $\beta$ are the strain coupling coefficients. Here, $\mathcal{A}$ simply modulates the energy difference between the ground and excited state and so is neglected. On the other hand, $\mathcal{E}_{x}$ and  $\mathcal{E}_{y}$ (which transform according to the $E_g$ irreducible representation of $D_{3d}$) have the effect of mixing the $\{\ket{e_{+}}, \ket{e_{-}}\}$ orbitals. We neglect DJT as it is typically weak. Upon the inclusion of spin to the basis; now written as $\{\ket{e_{+}}, \ket{e_{-}}\} \otimes \{\ket{\uparrow}, \ket{\downarrow}\}$, the spin-orbit interaction is described by the operator,
\begin{equation}
    \lambda L_z \otimes S_z,
\end{equation} where $\lambda$ is the spin-orbit coupling and $\{\sigma_i\ | i \in \{x, y, z\}\}$ are the Pauli matrices. In the ground state, $\lambda_g/2\pi \sim 165$~GHz whereas in the excited state, $\lambda_u/2\pi \sim 1098$~GHz~\cite{Maity2018}. The total Hamiltonian is,
\begin{equation}
    \label{eq::total_ham}
    H = \lambda L_z \otimes S_z + \begin{pmatrix} 0 & \alpha - i\beta \\ \alpha + i\beta & 0 \end{pmatrix} \otimes \mathbb{I}_2 + \mathbb{I}_2 \otimes (\gamma_e \mathbf{B} \cdot \mathbf{S})
\end{equation} where $\mathbb{I}_n$ is an n-dimensional identity operator, $\gamma_e \sim 28$~GHz/T is the electron gryomagnetic ratio, $\mathbf{B} = (B_x, B_y, B_z)^T$ is the applied magnetic field and $\mathbf{S} = (S_x, S_y, S_z)^T$ is the vector spin operator. The final term in Eq.~\ref{eq::total_ham} refers to the Zeeman effect.

\subsection{Perturbative Analysis of the Hamiltonian}

To better understand how strain affects the hyperfine structure and the optical selection rules of the nuclear spin, we perform a perturbative analysis of the GeV Hamiltonian. In this subsection, we begin with the electronic structure and in later subsections include the hyperfine structure. Here, we use canonical Van Vleck perturbation theory~\cite{Redmon2008} \red{to perturbatively reduce the four-dimensional Hamiltonian to an effective two-dimensional one via a transformation $e^S$}. To do so, we re-order the basis to, $\{\ket{e_{+}, \uparrow}, \ket{e_{-}, \downarrow}, \ket{e_{-}, \uparrow}, \ket{e_{+}, \downarrow}\}$. In this basis and in matrix form, the Hamiltonian may be written as,

\begin{equation}
    H = \frac{1}{2} 
    \begin{pmatrix} 
    \lambda + \gamma_e B_z & 0 & 2\epsilon & \gamma_e B_+  \\
    0 & \lambda - \gamma_e B_z & \gamma_e B_+^* & 2\epsilon^* \\
    2\epsilon^* & \gamma_e B_+ & -\lambda + \gamma_e B_z & 0 \\
    \gamma_e B_+^* & 2\epsilon & 0 & -\lambda - \gamma_e B_z  \\
    \end{pmatrix} 
\end{equation}  where $B_+ = B_x - iB_y$ and $\epsilon = \alpha - i\beta$. Note, we have neglected the orbital Zeeman effect as it is quenched. Assuming $\gamma_e |B_+|, \gamma_e B_z, |\epsilon| \ll \lambda$, we define two perturbation terms $V_{\rm{D}}$ and $V_{\rm{X}}$ and then partition $H$ such that,

\begin{equation}
    \label{eq::partition}
    H = \underbrace{
        \begin{pmatrix} 
            \lambda/2 & 0 & 0 & 0  \\
            0 & \lambda/2 & 0 & 0  \\
            0 & 0 & -\lambda/2 & 0 \\
            0 & 0 & 0 & -\lambda/2 \\
        \end{pmatrix}}_{H^{[0]}} + 
        \underbrace{
        \begin{pmatrix} 
            \gamma_e B_z/2 & 0 & 0 & 0    \\
            0 & - \gamma_e B_z/2 & 0 & 0  \\
            0 & 0 & \gamma_e B_z/2 & 0    \\
            0 & 0 & 0 & - \gamma_e B_z/2  \\
        \end{pmatrix}}_{V_{\rm{D}}} + 
        \underbrace{
        \begin{pmatrix} 
            0 & 0 & \epsilon & \gamma_e B_+/2  \\
            0 & 0 & \gamma_e B_+^*/2 & \epsilon^* \\
            \epsilon^* & \gamma_e B_+/2 & 0 & 0 \\
            \gamma_e B_+^*/2 & \epsilon & 0 & 0  \\
        \end{pmatrix}}_{V_{\rm{X}}}.
\end{equation} We compute the first-order expansion of $S$ by solving the commutator equation:
$[H^{[0]}, S^{[1]}] = -V_{\rm{X}}$ This is, 
\begin{equation}
    S^{[1]} = -\frac{1}{2\lambda}
    \begin{pmatrix} 
    0 & 0 & 2\epsilon & \gamma_e B_+  \\
    0 & 0 & \gamma_e B_+^* & 2\epsilon^* \\
    -2\epsilon^* & -\gamma_e B_+ & 0 & 0 \\
    -\gamma_e B_+^* & -2\epsilon & 0 & 0  \\
    \end{pmatrix}.
\end{equation}

We may then compute the first-order correction to $H$ as:
\begin{align}
    \label{eq::corrections}
    H^{[1]} &= V_{\rm{D}}, \\
    H^{[2]} &= \frac{1}{2} [V_{\rm{X}}, S^{[1]}]
\end{align} Therefore,
\begin{equation}
    H^{[2]} = \frac{1}{4\lambda}
    \begin{pmatrix} 
    4|\epsilon|^2 + \gamma_e^2 |B_+|^2 & 4\gamma_e B_+ \epsilon & 0 & 0     \\
    4\gamma_e B_+^* \epsilon^* & 4|\epsilon|^2 + \gamma_e^2 |B_+|^2 & 0 & 0     \\
    0 & 0 & -(4|\epsilon|^2 + \gamma_e^2 |B_+|^2) & -4\gamma_e B_+ \epsilon^*  \\
    0 & 0 & -4\gamma_e B_+^* \epsilon & -(4|\epsilon|^2 + \gamma_e^2 |B_+|^2)  \\
    \end{pmatrix}.
\end{equation} If restricted to the lowest energy branch, then - up to the second order - the effective Hamiltonian is,
\begin{equation}
    \label{eq::ham_eff}
    H^{\rm{eff}} = \mathbf{B} \mathbf{g} \mathbf{S},
\end{equation} where
\begin{equation}
    \mathbf{g} = \gamma_e
    \begin{pmatrix}
        -2 \alpha/\lambda & -2 \beta/\lambda & 0  \\
         2 \beta/\lambda & -2 \alpha/\lambda & 0  \\
        0 & 0 & 1
    \end{pmatrix}.
\end{equation} By diagonalizing $\mathbf{g}$, we find $g_\perp = 2\gamma_e |\epsilon| / \lambda$ and up to the first order, we may define $\ket{1(A)} = (\mathbb{I}_4 + S^{[1]}) \ket{e_+, \downarrow}$ and $\ket{2(B)} = (\mathbb{I}_4 + S^{[1]}) \ket{e_-, \uparrow}$, where $\mathbb{I}_4$ is a four-dimensional identity matrix.

\subsection{Optical Transition Dipoles}
\label{ssec:dipole}

The combined ground and excited state Hamiltonian is $H_{\rm{T}} = (H_u+E_0\mathbb{I}_4) \oplus H_g$, where $E_0/hc \sim 602$~nm is the zero phonon line (ZPL) energy and $\oplus$ is the matrix direct sum. Note, we have introduced the $g$ and $u$ subscripts to refer to ground and excited state. The dipole operators may be represented as the block matrices (denoted by $[\cdot]$),
\begin{equation}
    P_i = 
    \begin{bmatrix}
        0 & p_i \\
        p_i^\dag & 0 
    \end{bmatrix}
\end{equation} for $i \in \{x, y, z\}$ and $p_i$ is a $4$x$4$ matrix. Specifically, these are,
\begin{equation}
    p_x = 
    \begin{pmatrix}
        0 & 0 & 1 & 0 \\
        0 & 0 & 0 & 1 \\
        1 & 0 & 0 & 0 \\
        0 & 1 & 0 & 0 
    \end{pmatrix}, \,\,
    p_y = 
    \begin{pmatrix}
        0 & 0 & i & 0  \\
        0 & 0 & 0 & -i \\
        -i & 0 & 0 & 0 \\
        0 & i & 0 & 0 
    \end{pmatrix}, \,\, \textrm{and} \,\,
    p_z = 
    \begin{pmatrix}
        2 & 0 & 0 & 0  \\
        0 & 2 & 0 & 0 \\
        0 & 0 & 2 & 0 \\
        0 & 0 & 0 & 2 
    \end{pmatrix}.
\end{equation} Analogous to the construction of $H_T$, the first order expansion of $S$ on $H_T$ is $S_T^{[1]} = S_u^{[1]} \oplus S_g^{[1]}$. By applying the Baker-Campbell-Hausdorff relation we compute the effective dipole operators by evaluating,
\begin{equation}
    \label{eq::baker-campbell-hausdorff}
    e^{-S_T^{[1]}} P_i e^{S_T^{[1]}} \approx P_i + [P_i, S_T^{[1]}] + \frac{1}{2}[[P_i, S_T^{[1]}], S_T^{[1]}],
\end{equation} and collecting terms up to the second order. We introduce the parameters $\bar{\epsilon} = \epsilon/\lambda$, $\bar{B}_{+} = B_+/\lambda$ and $\tilde{p}_{i}$ as the two-dimensional matrix that couples the lower energy orbital branches of the ground and excited state. For example, $\tilde{p}_{z}^{[0]} \ket{e_{g,-}, \uparrow} = 2\ket{e_{u,-}, \uparrow}$ and $\tilde{p}_{z}^{[0]} \ket{e_{g,+}, \downarrow} = 2\ket{e_{u,+}, \downarrow}$. We compute the $1^{\rm{st}}$ order corrections to $\tilde{p}_{i}$ as,

\begin{equation}
\begin{split}
    \tilde{p}_{x}^{[1]} &= \frac{1}{2}
    \begin{pmatrix}
        -2(\bar\epsilon_g + \bar\epsilon_u^*) & -\gamma_e (\bar{B}_{+,g} + \bar{B}_{+,u}) \\
         -\gamma_e (\bar{B}_{+,g}^* + \bar{B}_{+,u}^*) & -2(\bar\epsilon_g^* + \bar\epsilon_u)
    \end{pmatrix} \approx -\frac{1}{2}
    \begin{pmatrix}
        2(\bar\epsilon_g + \bar\epsilon_u^*) & \gamma_e \bar{B}_{+,g} \\
         \gamma_e \bar{B}_{+,g}^* & 2(\bar\epsilon_g^* + \bar\epsilon_u)
    \end{pmatrix}, \\
    \\
    \tilde{p}_{y}^{[1]} &=  \frac{i}{2}
    \begin{pmatrix}
        2(\bar\epsilon_g - \bar\epsilon_u^*) & \gamma_e (\bar{B}_{+,g} + \bar{B}_{+,u}) \\
         -\gamma_e (\bar{B}_{+,g}^* + \bar{B}_{+,u}^*) & -2(\bar\epsilon_g^* - \bar\epsilon_u)
    \end{pmatrix} \approx \frac{i}{2}
    \begin{pmatrix}
        2(\bar\epsilon_g - \bar\epsilon_u^*) & \gamma_e \bar{B}_{+,g} \\
         -\gamma_e \bar{B}_{+,g}^* & -2(\bar\epsilon_g^* - \bar\epsilon_u)
    \end{pmatrix} \,\, \textrm{and} \\
    \\
    \tilde{p}_{z}^{[1]} &= 0.
\end{split} 
\end{equation} Here, we have made the approximation  $\gamma_e B_+ /\lambda_u \sim 0$. Similarly, the second order corrections are,
\begin{equation}
\begin{split}
    \tilde{p}_{x}^{[2]} &= 0, \\
    \tilde{p}_{y}^{[2]} &=  0 \,\, \textrm{and} \\
    \\
    \tilde{p}_{z}^{[2]} &\approx - \frac{1}{2}
    \begin{smallmatrix}
    \begin{pmatrix}
        \bar\epsilon_g (\bar\epsilon_g^* - \bar\epsilon_u^*) - \bar\epsilon_u^* (\bar\epsilon_g - \bar\epsilon_u) & 
        \gamma_e (\bar{B}_{+,g} - \bar{B}_{+,u}) (\bar\epsilon_g^* - \bar\epsilon_u^*) \\
        \gamma_e (\bar{B}_{+,g}^* - \bar{B}_{+,u}^*) (\bar\epsilon_g - \bar\epsilon_u) & 
        \bar\epsilon_g^* (\bar\epsilon_g - \bar\epsilon_u) - \bar\epsilon_u (\bar\epsilon_g^* - \bar\epsilon_u^*)
    \end{pmatrix}
    \end{smallmatrix},
\end{split}
\end{equation} where we have made the approximation $\gamma_e^2 |\bar{B}_{+,g}|^2 \approx \gamma_e^2 |\bar{B}_{+,u}|^2 \approx \gamma_e^2 |\bar{B}_{+,g} \bar{B}_{+,u}| \approx 0$.

\subsection{Hyperfine Structure}
\label{sec::hf_structure}

The hyperfine coupling adds to the Hamiltonian, the term, \red{
\begin{equation}
    H_{\rm{hf}} = \frac{A_\perp}{2} (S_- I_+ + S_+ I_-) + A_\parallel S_z I_z,
\end{equation} 
where $A_\perp$ and $A_\parallel$ are the transverse and longitudinal hyperfine interaction strengths,  $S_{\pm} = S_x \pm iS_y$ and $I_{\pm} = I_x \pm iI_y$} and $I_x$, $I_y$ and $I_z$ are the nuclear spin operators. For the ordered basis $\{\ket{e_{+}, \uparrow}, \ket{e_{-}, \downarrow}, \ket{e_{-}, \uparrow}, \ket{e_{+}, \downarrow}\} \otimes \{\ket{m} |\, m \in \{-I, -I+1, ..., I\}\}$, we can partition the Hamiltonian as done in~\ref{eq::partition}. More explicitly, this is,

\begin{equation}
    H_{\rm{el-n}} = \underbrace{H^{[0]} \otimes \mathbb{I}_{\rm{10}}}_{H_{\rm{el-n}}^{[0]}} +
                    \underbrace{V_{\rm{D}} \otimes \mathbb{I}_{10} + A_\parallel \mathbb{I}_2 \otimes S_z \otimes I_z}_{V_{\rm{D,el-n}}} +
                    \underbrace{
                    \frac{1}{2} \begin{bmatrix} 
                        0 & 0 & 2\epsilon \mathbb{I}_{10} & \gamma_e B_+ \mathbb{I}_{10} + A_\perp I_{-} \\
                        0 & 0 & \gamma_e B_+^* \mathbb{I}_{10} + A_\perp I_{+} & 2\epsilon^* \mathbb{I}_{10} \\
                        2\epsilon^* \mathbb{I}_{10} & \gamma_e B_+ \mathbb{I}_{10} + A_\perp I_{-} & 0 & 0 \\
                        \gamma_e B_+^* \mathbb{I}_{10} + A_\perp I_+ & 2\epsilon \mathbb{I}_{10} & 0 & 0  \\
                    \end{bmatrix}}_{V_{\rm{X,el-n}}}.
\end{equation} We determine 
\begin{equation}
    S^{[1]}_{\rm{el-n}} = -\frac{1}{2\lambda}
    \begin{bmatrix} 
    0 & 0 & 2\epsilon\mathbb{I}_{10} & \gamma_e B_+ \mathbb{I}_{10}  + A_\perp I_{-}\\
    0 & 0 & \gamma_e B_+^* \mathbb{I}_{10} + A_\perp I_{+} & 2\epsilon^* \mathbb{I}_{10} \\
    -2\epsilon^* \mathbb{I}_{10} & -\gamma_e B_+ \mathbb{I}_{10} -  A_\perp I_{-} & 0 & 0 \\
    -\gamma_e B_+^* \mathbb{I}_{10} - A_\perp I_{+} & -2\epsilon \mathbb{I}_{10} & 0 & 0  \\
    \end{bmatrix}
\end{equation} and similarly, from Eq.~\ref{eq::corrections},
\begin{equation}
    \begin{split}
    H^{[2]}_{\rm{el-n}} &= H^{[2]} \otimes \mathbb{I}_{10} + \\
    \\
    \frac{1}{4\lambda} &
    \begin{bmatrix}
    \begin{smallmatrix} 
    2 \gamma_e A_\perp (B_x I_x + B_y I_y) + A_\perp^2 I_{-}I_+ &  4\epsilon A I_{-} & 0 & 0 \\
    4\epsilon^* A_\perp I_+ & 2 \gamma_e A (B_x I_x + B_y I_y) + A_\perp^2 I_+I_{-} & 0 & 0 \\
    0 & 0 & -(2 \gamma_e A_\perp (B_x I_x + B_y I_y) + A_\perp^2 I_{-}I_+) & -4\epsilon^* A_\perp I_{-}\\
    0 & 0 & -4\epsilon A_\perp I_{+} & -(2 \gamma_e A_\perp (B_x I_x + B_y I_y) + A_\perp^2 I_{+}I_{-})  \\
    \end{smallmatrix}        
    \end{bmatrix}.
    \end{split}
\end{equation} If we restrict ourselves to the lowest energy orbital branch, then - up to the second order - the effective Hamiltonian is,
\begin{equation}
    {H}^{\rm{eff}}_{\rm{el-n}} = \mathbf{B} \mathbf{g} \mathbf{S} + \mathbf{B} \mathbf{g}_n \mathbf{I} + \mathbf{S} \mathbf{A} \mathbf{I},
\end{equation} where,
\begin{equation}
    \mathbf{g}_n = \frac{\gamma_e A_\perp}{2\lambda}
    \begin{pmatrix}
        -1 & 0 & 0 \\
        0 & -1 & 0 \\
        0 & 0 & 0
    \end{pmatrix} \,\, \textrm{and} \,\,
    \mathbf{A} = 
    \begin{pmatrix}
        -4\alpha A_\perp/\lambda & -4\beta A_\perp/\lambda & 0 \\
        4\beta A_\perp/\lambda & -4\alpha A_\perp/\lambda & 0 \\
        0 & 0 & A_\parallel
    \end{pmatrix}.
\end{equation}
Here, we have neglected the second order hyperfine; the terms $A_\perp^2 I_{-} I_+/4\lambda$ and $A_\perp^2 I_{+} I_-/4\lambda$, as these terms have an interaction strength, $A_\perp^2/8\pi\lambda \sim 2$~kHz in the ground state. We compute the effective nuclear gryomagnetic ratio and hyperfine coupling as $g_{n, \perp} = \gamma_e A_\perp/2\lambda \sim 3.1$~MHz/T and $A^{\rm{eff}}_{\perp} = 4|\epsilon|A_\perp/\lambda$. \red{Note, that the transverse hyperfine takes on a strain dependence}. In our analysis, we have ignored the nuclear Zeeman effect due to the weak \Ge{73} gyromagnetic ratio of $\gamma_n \sim 1.484$~MHz/T. The low nuclear gyromagnetic ratio makes coherent control of the \Ge{73} via nuclear magnetic resonance difficult, particularly in the environment of a dilution refrigerator with limited cooling power. However, as $g_{n,\perp} > \gamma_n$ this issue should be somewhat alleviated. Analogous to the electron transverse g-factor, the effective transverse hyperfine is strain dependent and therefore, in a similar vein to the acoustic drive of the electron spin \cite{Maity2022}, so too should the acoustic drive of an electron-nuclear flip-flop transition be possible. The  strain susceptibility of the transition $4 A_\perp d/\lambda \sim 2.5$~THz/strain, where $d \sim 2.8$~PHz/strain is the GeV strain coupling factor~\cite{Maity2018} (for the $XX-YY$ component of the strain tensor).

We may extend the analysis performed on the dipole operators in Sec.~\ref{ssec:dipole} to include the hyperfine structure. With the addition of the nuclear spin, the dipole operators become $P_{i,\rm{el-n}} = P_{i} \otimes \mathbb{I}_{10}$. We introduce the parameter $\bar{A} = A_\perp/\lambda$ and follow the procedure in Eq.~\ref{eq::baker-campbell-hausdorff} to compute the $1^{\rm{st}}$ and $2^{\rm{nd}}$ order corrections to the dipole matrices as,
\begin{equation}
\begin{split}
    \tilde{p}_{x, \rm{el-n}}^{[1]} &=  \tilde{p}_{x}^{[1]} \otimes \mathbb{I}_{10} - \frac{\bar{A}_u + \bar{A}_g}{2}
    \begin{bmatrix}
        0 & I_{-}\\
        I_{+} & 0
    \end{bmatrix}, \\
    \\
    \tilde{p}_{y, \rm{el-n}}^{[1]} &=  \tilde{p}_{y}^{[1]} \otimes \mathbb{I}_{10} + i\frac{\bar{A}_u + \bar{A}_g}{2}
    \begin{bmatrix}
        0 & I_{-}\\
        -I_{+} & 0
    \end{bmatrix}, \\
    \tilde{p}_{z, \rm{el-n}}^{[1]} &= 0, \\
    \\
    \tilde{p}_{x, \rm{el-n}}^{[2]} &=  0, \\
    \\
    \tilde{p}_{y, \rm{el-n}}^{[2]} &=  0, \,\, \textrm{and} \\
    \tilde{p}_{z, \rm{el-n}}^{[2]} &\approx \tilde{p}_{z}^{[1]} \otimes \mathbb{I}_{10} + (\bar{A}_u - \bar{A}_g)
    \begin{bmatrix}
        0 & (\bar{\epsilon_g}^* - \bar{\epsilon_u}^*) I_{-}\\
        (\bar{\epsilon_g} - \bar{\epsilon_u}) I_{+} & 0
    \end{bmatrix}.
\end{split}
\end{equation}
\red{Note, the transverse hyperfine produces nuclear spin non-conserving optical transitions and the strength of these transitions is modulated by the difference between the ground- and excited-state strain.}

\section{Strain Dependence of the Hyperfine Splitting}

\begin{figure}
\centering
\includegraphics{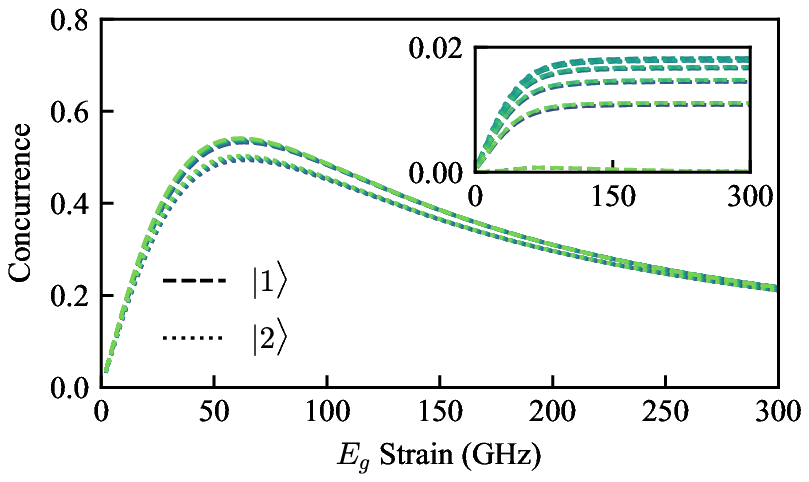}
\caption{\label{fig:concurrence} Concurrence of the GeV eigenstates computed (main panel) from the orbital states with respect to the electron-nuclear system and (inset) from the orbital-electron states with respect to the nuclear system.}
\end{figure}

In Fig.~\ref{fig:concurrence}, we plot the strain dependence the concurrence~\cite{Rungta2001} of the GeV eigenstates. Here, concurrence is used as a measure of separability of the GeV orbital and electron-nuclear spin degrees of freedom. The hyperfine splitting (see Main Text) and concurrence are correlated and allude to the following; as strain is increased the orbital and electron spin mix and the electron spin becomes less defined. Consequently, the hyperfine splitting decreases. However, when strain is further increased, and the system approaches maximal orbital-spin mixing, the process reverses; the electron spin is once again well defined and the hyperfine splitting approaches $A_\parallel$.

\section{Sample Preparation}
\label{sec::sample_preparation}

The sample used in this work is a $50$~$\mu$m thick, electronic grade diamond membrane (Element Six) with a $\langle100\rangle$ oriented surface. We implant \Ge{73} at a fluence of $2.5 \times 10^{11}$~Ge/cm\textsuperscript{2} and at an energy of $220$~keV. From stopping and range of ions in matter (SRIM) simulations, we approximate the implantation depth to be $\sim 83$~nm. Following the implantation, \red{the sample is annealed at $800^\circ$C} for $10$~hrs. Finally, the sample is cleaned using a Bristol boil (a mixture of sulphuric acid and sodium nitrate) at \red{$400^\circ$C} for $30$~min to remove any graphitization that occurred during the previous steps.

\section{Experimental Setup}

\begin{figure}
\centering
\includegraphics{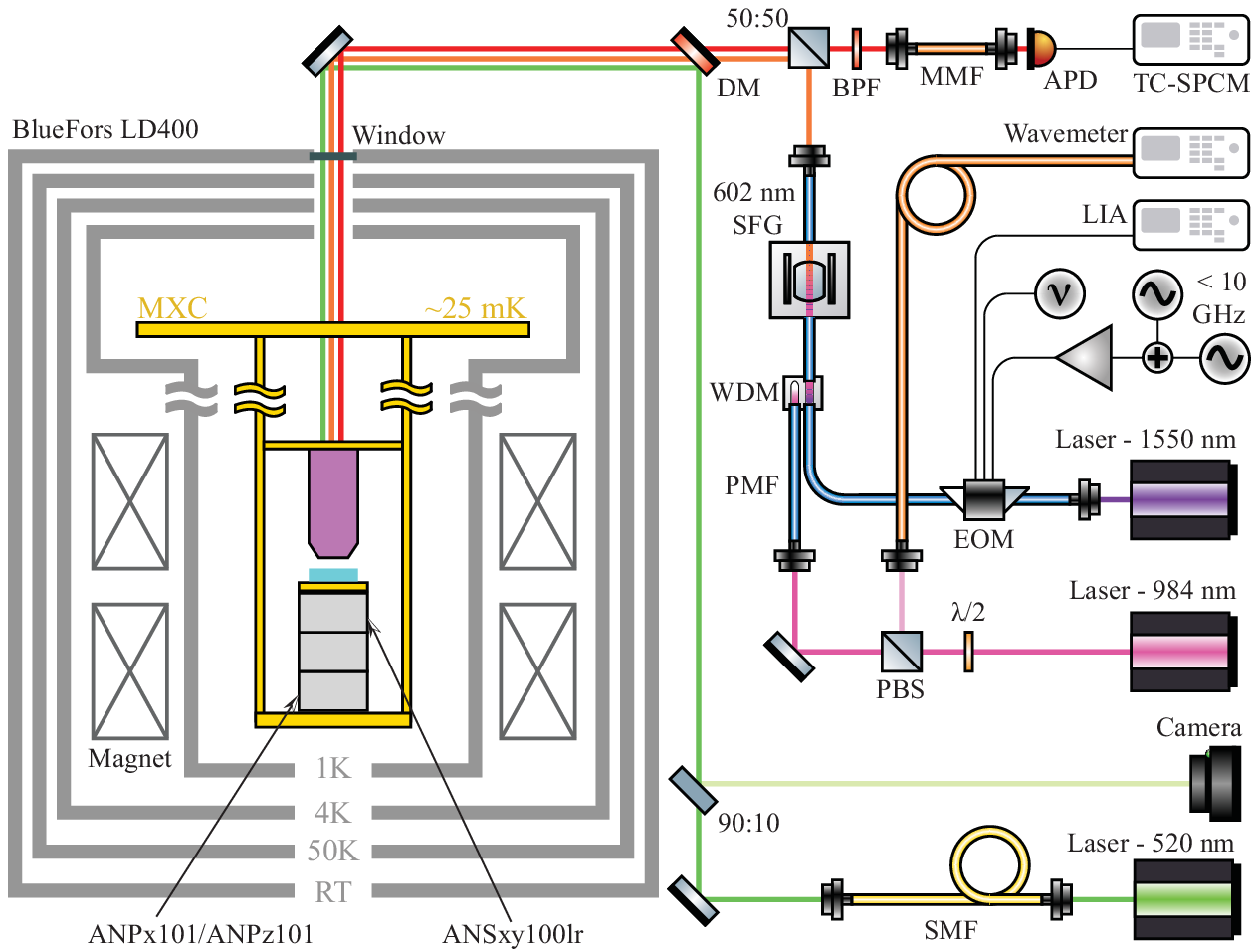}
\caption{\label{fig:fridge} Schematic diagram of the experimental setup.}
\end{figure}

The experiments described in this work were performed with a home-built confocal microscope (see Fig.~\ref{fig:fridge}). The sample is mounted onto nano-positioner stack consisting of two Attocube Systems AG ANPx101 and an ANPz101 that provide three-axis coarse ($\sim 1$~$\mathrm{\mu}$m) positioning and an ANSxy100lr that offers fine ($< 100$~nm) positioning capabilities in the $xy$-plane. The sample and positioner stack are mounted at the end of a cold-finger thermally connected to the mixing chamber (MXC) flange of a BlueFors Oy. LD400 dilution refrigerator. The data plotted in Fig.~\ref{fig:g2}-\ref{fig:zeeman} was measured at $4$~K whilst the data in the remainder of this work was measured at mK. \red{Operation at mK is necessary to prevent decoherence that results from thermal excitation into the higher orbital branch that is at least $\sim 165$~GHz above the ground state spin sub-levels.} The excitation lasers are focused onto the sample by a microscope objective lens ($\mathrm{NA} = 0.87$, $\mathrm{WD} = 0.9$~mm) that is also mounted to the the cold-finger. A $1$~mm diameter, hemispherical zirconia solid immersion lens ($n = 2.15$) is glued onto the sample with vacuum grease (Dow Corning) to improve PLE collection efficiency.

Off-resonant excitation is provided by a $520$~nm diode laser (Thorlabs Inc. LP520-SF15A) with its peak power set to $\sim 1$~$\mathrm{\mu}$W. Pulses are generated by modulation of the diode current. The GeV has a ZPL at $\sim 602$~nm. For resonant excitation at this wavelength, we use a nonlinear crystal for sum-frequency generation (SFG) of a $1550$~nm fixed-frequency external cavity laser (Thorlabs Inc. SFL1550P) and a $984$~nm tuneable external cavity laser (Toptica AG DLPro). The two infrared lasers are combined in a fiber wavelength division multiplexer (WDM) prior to the SFG module (AdvR Inc.). An electro-optic modulator (EOM, iXblue SAS) is used for amplitude modulation of the $1550$~nm laser. A DC voltage biases the EOM to the carrier suppressed regime. This voltage is determined by measuring the EOM carrier output with a built-in photodiode and a lock-in amplifier (LIA, Stanford Research Systems SR830) and then feeding this signal into a \textit{proportional-integral-derivative} (PID) controller implemented on the measurement computer. A carrier extinction ratio of $30$~dB is achieved. The EOM is driven by a Keysight Technologies N5183B and an Agilent Technologies Inc. E4438C microwave source. The outputs of which are combined and then amplified before the EOM. \red{Microwave drive of the EOM in the carrier suppressed regime produces two equal power sidebands, separated by twice the microwave frequency, that we use to perform two laser spectroscopy and coherent population trapping (CPT)}. Resonant pulses are generated by pulse modulation of the microwave sources. Pulse generation is controlled by a SpinCore Technologies Inc. PulseBlasterESR-Pro. To stabilize the laser wavelength, the $984$~nm laser is sampled by a half-wave plate ($\lambda/2$) and polarizing beam splitter (PBS) and then fed into a High Finesse GmbH WS-7 wavemeter that measures its wavelength. A digitally implemented PID controller then stabilizes the wavelength from the error signal produced by the wavemeter.  The resonant excitation laser is multiplexed with its off-resonant counterpart in free-space with a $587$~nm long-pass dichroic mirror (DM) prior to being sent into the cryostat. The single mode (SMF) and polarization maintaining (PMF) fibers that launch the excitation lasers ensure that the beams exhibit a Gaussian mode profile.

Photoluminescence excitation (PLE) is collected by the microscope objective and subsequently filtered, both spectrally [BPF, $630-670$~nm passband in the GeV phonon sideband (PSB)] and spatially (MMF, core size $\sim 110$~$\mathrm{\mu}$m), prior to detection by an avalanche photodiode (APD, Excelitas Technologies Inc. SPCM-AQRH). In response to incident single photons, the APD generates low-voltage transistor-transistor logic (LVTTL) pulses that are counted by a time-correlated single single photon counting module (TC-SPCM, qutools GmbH QuTag).

\section{Defect Characterization}
\label{sec::defect_char}

\begin{figure}
\centering
\includegraphics{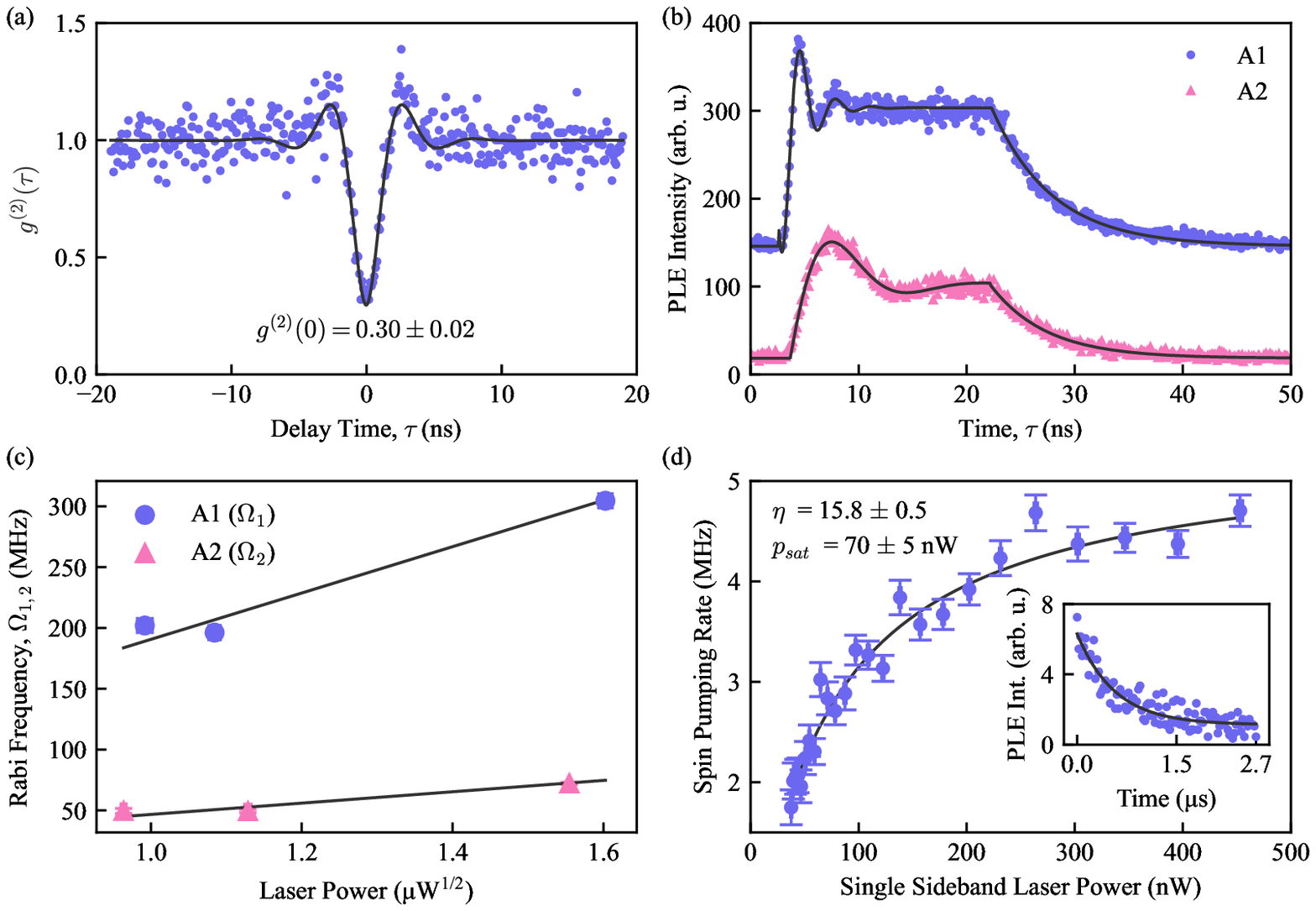}
\caption{\label{fig:g2} 
(a) Second order autocorrelation of the PLE emission from the GeV used in this work. 
(b) Time-resolved histogram of the GeV photoluminescence during which a $20$~ns laser pulse is applied to the A1 (blue circles) and A2 (purple triangles) transitions. The data is fit (black line) to a decaying sinusoid for the duration of the pulse followed by an exponential decay. 
(c) Relationship between the A1 and A2 optical Rabi frequencies and excitation power. 
(d) Optical spin pumping rate as a function of excitation power. (inset) Optical spin pumping; driving the A1 transition results in an exponential decay in the PLE intensity as the spin is pumped into $\ket{2}$.}
\end{figure}

In Fig.~\ref{fig:g2}(a) we plot the second order autocorrelation, $g^{(2)}(\tau)$ of the GeV PLE. To measure this quantity, we construct a Hanbury-Brown-Twiss interferometer by replacing the MMF fiber on the detection arm of the confocal microscope (see Fig.~\ref{fig:fridge}) with a fiber optic coupler with two APDs connected to the two outputs and measuring coincident photons. The experiment is performed under resonant excitation and consequently we fit the data to \cite{Chen2022}

\begin{equation}
    g^{(2)}(\tau) = 1- \beta e^{-\zeta\tau} \bigg[\cos{(\nu\tau)} + \frac{\zeta}{\nu} \sin{(\nu\tau)}\bigg],
\end{equation} where $\zeta = (1/\tau_r + 1/\tau_2)/2$ and  $\nu = \sqrt{\Omega^2 - (1/\tau_r - 1/\tau_2)^2/4}$, and where $\tau_{\mathrm{r}}$, $\tau_2$ and $\Omega$ are the optical lifetime, dephasing time and the Rabi frequency, respectively. We find $g^{(2)}(0) = 0.30 \pm 0.02 < 0.5$, confirming that the emitted light is indeed from a single GeV colour centre.

\begin{figure}
\centering
\includegraphics{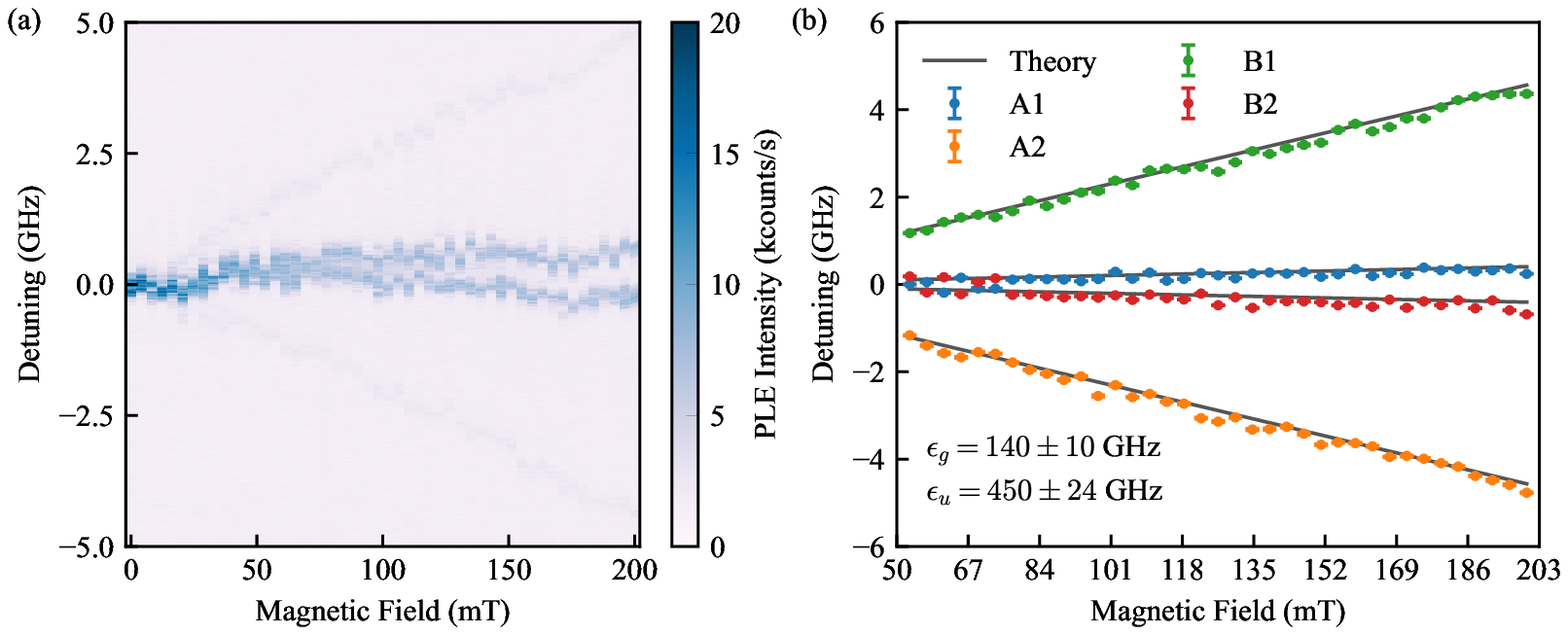}
\caption{\label{fig:zeeman} (a) Magnetic field dependence of the transition of the GeV used in this work. (b) The relative resonance frequencies of the transitions shown in (a). The resonances are fit (grey) to the GeV Hamiltonian. A data smoothing post-processing step is applied to the data in (a) prior to the fit, to reduce the spectral diffusion resultant error. From the fit, we extract the ground and excited state $E_g$-type strain splittings, here referred to  as $\epsilon_g$ and $\epsilon_u$.}
\end{figure}

We apply a magnetic field of $B_0 = 0.36$~T to split the ZPL into two \textit{spin-conserving} transitions, A1 and B2 and two \textit{spin-flipping} transitions, A2 and B1 (compare also Main Text). By exciting A1 or A2 with a $20$~ns resonant pulse and measuring the PLE intensity, we measure optical Rabi oscillations as shown in Fig.~\ref{fig:g2}(b). The linear relationship between Rabi frequency and the root of the excitation power confirms that these are Rabi oscillations. From the Rabi frequency, we extract a branching ratio of $\eta \approx \Omega_1^2/\Omega_2^2 = 16 \pm 5$. The exponential decay in PLE intensity after the pulse is switched \textit{off} corresponds to the optical lifetime. We find this to be $\tau_{\rm{r}} = 5.9 \pm 0.2$~ns. If a longer pulse, on the order of several $\sim \rm\mu$s, is used to drive the A1 transition, the system is then pumped into $\ket{2}$ due to optical relaxation via A2. Consequently, an exponential decay in the PLE intensity is observed as shown in the inset of Fig.~\ref{fig:g2}(d). In Fig.~\ref{fig:g2}(d), we plot the A1 spin pumping rate as a function of excitation power, $p$ and fit it to $pT_{\rm{P}}^{-1}/(p + p_{\rm{sat}})$ where $p_{\rm{sat}} = 70 \pm 5$~nW is the saturation power and $T_{\rm{P}}^{-1} = 5.4 \pm 0.1$~MHz is the saturation pumping rate. We note that $T_{\rm{P}} \approx 2\eta\tau_{\rm{r}}$~\cite{Debroux2021} and compute a branching ratio of $\eta = 15.8 \pm 0.5$. 


To optically observe the Zeeman effect, we perform PLE spectroscopy and sweep the magnetic field as shown in Fig.~\ref{fig:zeeman}(a). We fit the data to the GeV Hamiltonian [see Fig.~\ref{fig:zeeman}(b)] from which we extract the ground and excited state strain splittings. We find these to be $\epsilon_g = 140 \pm 10$~GHz and $\epsilon_u = 450 \pm 24$~GHz, respectively. We also extract $\epsilon_g = 141 \pm 2$~GHz by measuring the Zeeman splitting with CPT as shown in Fig. 2(a) of the main text. 

\begin{figure}[t]
\centering
\includegraphics{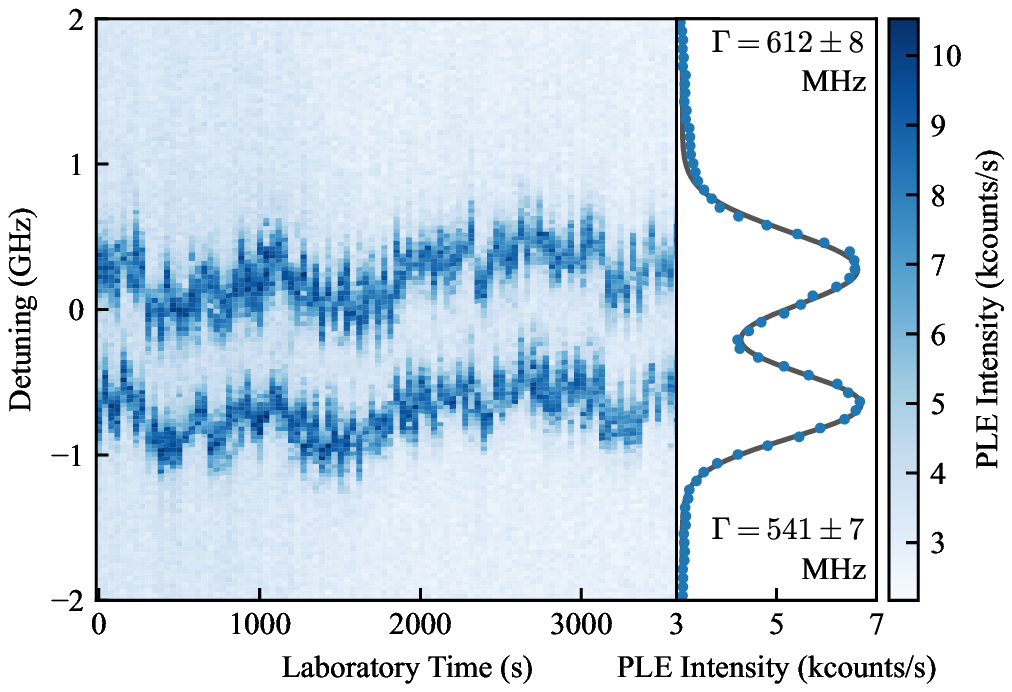}
\caption{\label{fig:specdiff} (left) PLE spectrum of the GeV used in this work, measured over the course of $\sim 1$~hr. (right) The mean PLE intensity measured over the course of the aforementioned hour. The data (blue dots) has been fit (black line) to a pair of Gaussian peaks. The full width half maximum ($\Gamma$) of each peak is shown adjacent to that peak.}
\end{figure}

Finally, we perform PLE spectroscopy over the course of an hour as shown in Fig. \ref{fig:specdiff}. We observe spectral fluctuation on the order of $\sim 100$ MHz with the long-term linewidth of the emitter being $\sim 600$~MHz. \red{This linewidth is beyond the lifetime limit and is due to inhomogeneous broadening.} A magnetic field of $200$~mT is applied; the two peaks shown are the A1 and B2 transitions. It should be noted that although significant spectral diffusion is present, we are able to resolve the spin-selective optical transitions.

\section{Modelling Coherent Population Trapping}

To perform CPT, two lasers of frequency $\omega_1$ and $\omega_2$ drive the A1 and A2 transitions with Rabi frequencies $\Omega_1$ and $\Omega_2$, respectively. \red{We define the one and two laser detunings as $\Delta = (\omega_1 + \omega_2)/2 - \omega_o$ and $\delta = (\omega_1 - \omega_2)-\omega_z$ for an optical splitting of $\omega_0$ and a Zeeman splitting of $\omega_z$. Note, that modifying the frequency difference between the two lasers - for example, by changing the frequency of the microwaves applied to the EOM (see Sec.~\ref{sec::sample_preparation}) - changes $\delta$ without changing $\Delta$.} We describe CPT - in the rotating frame - with the  Hamiltonian,

\begin{equation}
    H/\hbar = \delta \ketbra{\rm 2}{\rm 2} + \Delta \ketbra{\rm A}{\rm A} + \frac{\Omega_1}{2} \big(\ketbra{1}{\rm A} + \rm{h.c.} \big) + \frac{\Omega_2}{2}\big(\ketbra{2}{\rm A} + \rm{h.c.} \big).
\end{equation}

 It can be shown \cite{Agapev1993}, that under resonance i.e., $\Delta = \delta = 0$, one of the eigenstates of $H$ does not contain the excited state $\ket{\rm A}$ (this eigenstate is referred to as the dark state, $\ket{D}$) and is decoupled from the driving fields. Conversely, the two other eigenstates (referred to as the bright states, $\ket{B_\pm}$) both contain $\ket{\rm A}$ and can therefore relax into $\ket{D}$; emitting a photon as they do so. Upon photon emission, the system is trapped in $\ket{D}$. It should be noted that this only occurs when the one and two laser resonance conditions are met. If $\delta \neq 0$ and $\Delta = 0$, all the eigenstates of $H$ contain $\ket{\rm A}$ and are all bright.

 As a dissipative process, we model CPT with the master equation \cite{Manzano2020},

\begin{equation}
    \frac{d\rho}{dt} = \mathcal{L} [\rho] =  -\frac{i}{\hbar}[H, \rho] + \sum_{n}\big(C_n \rho C_n^\dag - \frac{1}{2}\{C_n^\dag C_n, \rho \}\big),
\end{equation}

where $\rho$ is the system density matrix and $\{C_n\}$ are a set of operators representing dissipation \red{and decoherence} in the system. \red{For the GeV, we model the radiative decay between $\ket{\rm A}$ and $\ket{1}$ and $\ket{2}$~\cite{Debroux2021} and the transverse decay between $\ket{1}$ and $\ket{2}$~\cite{Carter2021} with the operators:} 

\begin{equation}
    C_{\rm A1} = \sqrt{\gamma_{\rm A1}} \ket{1} \! \! \bra{\rm A},\,\, 
    C_{\rm A2} = \sqrt{\gamma_{\rm A2}} \ket{2} \! \! \bra{\rm A},\,\,
    C_\phi = \sqrt{\frac{\gamma_\phi}{2}} (\ket{2} \! \! \bra{2} - \ket{1} \! \! \bra{1}).
\end{equation}

Here, $\gamma_{\rm A1}$ and $\gamma_{\rm A2}$ are the radiative decay rates through A1 and A2. Note, $\gamma_{\rm A1} + \gamma_{\rm A2} = \tau_{\rm r}^{-1}$ and $\gamma_{\rm A1}/\gamma_{\rm A2} = \eta$. On the other hand, $\gamma_\phi$ is the \red{dephasing} rate. We assume that dephasing is the dominant with regards to $T_2^*$ as is typical for solid state spins. 

\begin{figure}
\centering
\includegraphics{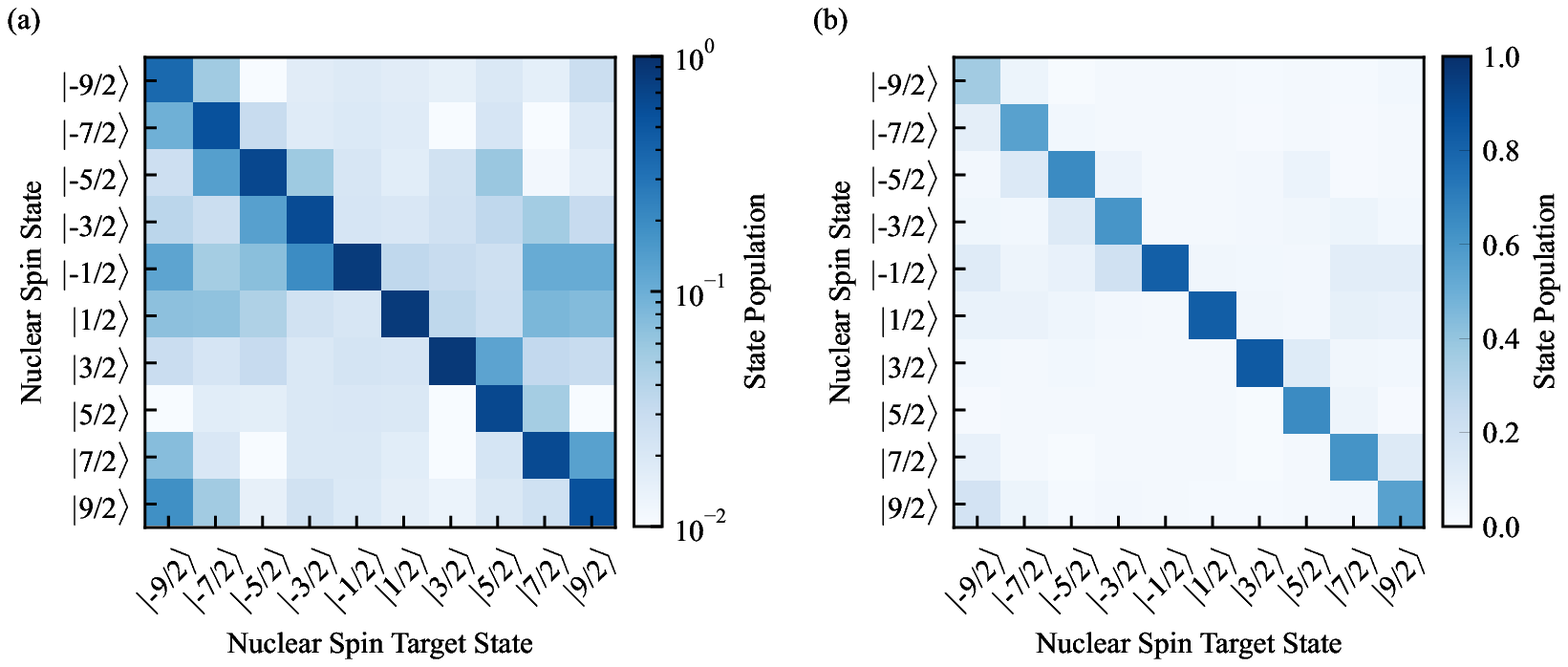}
\caption{\label{fig:init_probs} The measured nuclear spin state populations when attempting to initialize into each specified target state plotted with a logarithmic (a) and linear (b) scale. The population is computed from the data plotted Fig. 3(e) of the main text.}
\end{figure}

\red{As the CPT spectrum probes the steady state of the system, the CPT curve can be modelled by solving $\mathcal{L} [\rho_{\rm ss}] = 0$ and then evaluating ${\rm Tr}(\ket{\rm A} \!\! \bra{\rm A}\rho_{\rm ss})$. The shape of the CPT curve depends on the optical Rabi frequencies, the optical lifetime, branching ratio and the $T_2^*$ which for the GeV in similar operating conditions has been measured to be $T_2^* \sim 220$~ns~\cite{Adambukulam2023b}. We can represent this with the function ${\rm CPT}(\delta; \Delta, \Omega_1, \Omega_2, \tau_{\rm r}, \eta, T_2^*)$ that solves for ${\rm Tr}(\ket{\rm A} \!\! \bra{\rm A}\rho_{\rm ss})$ for a particular set of experimental parameters. Given the characterization measurements performed in Sec.~\ref{sec::defect_char}, we fix these experimental parameters such that the CPT curve is given by the function $ f(\delta, \Omega_1) = {\rm CPT}(\delta,; \Delta = 0, \Omega_1, \Omega_2 = \Omega_1/\eta, \tau_{\rm r} = 5.9 \times 10^{-9}, \eta = 15.8, T_2^* = 200 \times 10^{-9})$. Here, we have specified the arguments of $f$ in SI units. The data in Fig.~3(e) of the Main Text was fit to the model $A f(\delta - \delta_0, \Omega_1) + C$ where $A$ and $C$ are scale and offset parameters that account for the experimental data being measured in photons rather than the state population of $\ket{A}$ and $\delta_0$ models the nuclear state dependent shift in the resonance. To evaluate the initialization fidelity, we now fit each curve in Fig.~3(e) of the Main Text with the model,
\begin{equation}
    \sum_m |A_m| f(\delta - \delta_0^{m}, \Omega_1) + C
    \label{eq::fidelity_fit_model}
\end{equation} 
where $A_m$ and $\delta_0^m$ are the scale factor and \textit{dip} frequency for a nuclear state $\ket{m}$. The only parameters varied during the fit are the set of $A_m$ and $C$ as well as $\Omega_1$. We assume that the set of $A_m$ are normally distributed random variables with mean $\braket{A_m}$ and standard deviation $\Delta A_m$ that are extracted from the fit. We then perform a Monte-Carlo simulation where we collect $5 \times 10^{4}$ samples from the aforementioned normal distributions and for the $i^{\rm th}$ sample compute the state population of $\ket{n}$ as $P^i_n = |A^i_n|/(\sum_m |A^i_m|)$. The mean of these sampled populations, $\braket{P_{n}}$, is plotted in Fig.~\ref{fig:init_probs}. The initialization fidelity is $\braket{P_{\rm target}}$. Readout fidelity may be inferred from the error in initialization fidelity - here, given as the $95\%$ confidence interval as computed from the Monte-Carlo simulation.} 

\red{The use of Eq.~\ref{eq::fidelity_fit_model} is valid under the assumption that the CPT curve was measured before nuclear spin diffusion takes place. This is not the case when measuring continuous wave CPT as was done in Fig.~2(a) of the Main Text. Instead, we note that a CPT curve is the sum of a narrow, negative-valued Lorentzian (referred to as the \textit{dip}) and a broader, positive-valued \textit{peak}. Given the difference in linewidth of the \textit{dip} and \textit{peak}, in the vicinity of the \textit{dip}, it is sufficient to model the \textit{peak} up to its first derivative. Thus, we use the model, $A/(1-(\frac{\delta-\delta_0}{\gamma})^2) + C(\delta-\delta_0) + D$. }

\section{Nuclear Spin Diffusion Rate Model}

\begin{figure}
\centering
\includegraphics{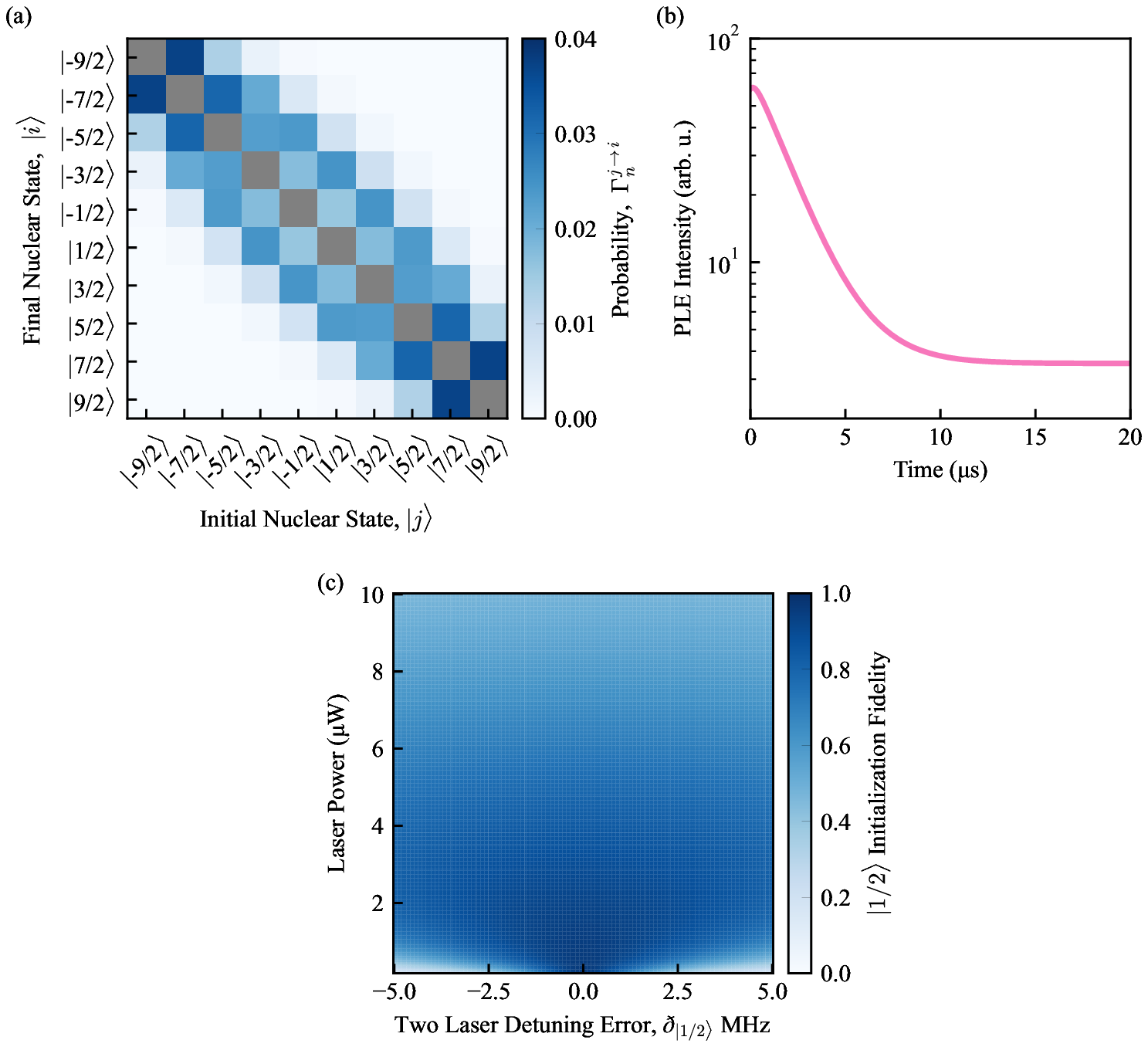}
\caption{\label{fig:spin_flip} (a) The calculated probability of nuclear spin flip given the parameters specified in the parameters for our experimental conditions assuming the ground state hyperfine coupling is isotropic and that in the excited state, $A_{u,\parallel} = 19$~MHz and $A_{u, \perp} = 10$~MHz. (b) Simulated PLE intensity as function of time during the pumping of $\ket{{1/2}}$. (c) Initialization fidelity for $\ket{1/2}$ as function of excitation power and error in the two laser detuning. The data is generated from the rate model, using the parameters specified in the text.}
\end{figure}

Given the large separation in timescales between the dynamics of the electron ($10$s of ns) and that of the nucleus (several $\rm{\mu}$s), we may assume that electron is always in the steady state. This simplifies the modelling as we may discard the coherent dynamics of the system and rather than solve the master equation, instead solve a rate model. In its most general form, the rate model is expressed as,
\begin{equation}
    \frac{dP_i}{dt} = \frac{1}{2} \gamma_r \bigg[\mathbb{P}_{\rm{exc}}^{j} \mathbb{P}^{j \rightarrow i}_n P_j - \delta_{ij} \bigg( \sum_{l \neq i}  \mathbb{P}_{\rm{exc}}^{l} \mathbb{P}^{i \rightarrow l}_n \bigg) P_i\bigg],
\end{equation} where $P_j$ is the population in the nuclear spin state $\ket{j}$; where $j \in \{-I, -I+1, ..., I\}$, $\mathbb{P}_{\rm{exc}}^j$ is the excitation probability, $\mathbb{P}^{j \rightarrow i}_n$ is the probability the nuclear spin flips from $\ket{j}$ to $\ket{i}$ and $\delta_{ij}$ is the Kronecker delta. The shape of a CPT spectrum implies the probability of optical excitation is the sum of two Lorentzian distributions, the second of which corresponds to the CPT \textit{dip}; has a depth $\propto 1-\tau_{\rm{r}}/2 T_2^{*}$ and has the effect of suppressing optical excitation. We also note that the transition saturates, as seen in Fig.~\ref{fig:g2}(d). To capture these phenomena, we define the optical excitation probability as
\begin{equation}
    \mathbb{P}_{\rm{exc}}^{j} =
    \underbrace{\frac{p}{p+p_{\rm{sat}}}}_{\rm{Saturation}}
        \bigg[\underbrace{\frac{1}{1+(\eth_j/\Gamma)^2}}_{\rm{Excitation}} - 
        \underbrace{\bigg( 1 - \frac{\tau_{\rm{r}}}{2 T_2^*} \bigg) \frac{1}{1 + (\eth_j/\gamma)^2}}_{\rm{CPT \,\, \textit{dip}}} \bigg],
\end{equation} where $\eth_j = \delta - \delta_j$ is the difference between the applied two-laser detuning, $\delta$ and the two-laser detuning, $\delta_j$, resonant to $\ket{j}$, $p$ is the laser power and $\Gamma = \sqrt{\Gamma_0^2 + \Omega_1^2/4\pi^2 + \Omega_2^2/4\pi^2}$ and $\gamma = \sqrt{\gamma_0^2 + (\Omega_1 \Omega_2 /4\pi^2 \gamma_{\rm{r}} )^2}$ are the optical and CPT \textit{dip} linewidths when the natural optical and CPT linewidths are $\Gamma_0$ and $\gamma_0 = 1/2\pi T_2^*$. We now turn our attention to $\mathbb{P}^{i \rightarrow j}_n$. From Fermi's golden rule and for $\ket{\mu} \in \{\ket{1}, \ket{2}\}$, the transition probability from $\ket{A, j}$ to $\ket{\mu, i}$ is 
\begin{equation}
    \mathbb{P}^{A,j \rightarrow \mu, i} \propto \mathcal{P}^{A,j \rightarrow \mu, i} = \sum_{k}  |\!\bra{A, i} P_{k, \rm{el-n}} \ket{\mu, i}\!|^2, 
\end{equation} where $P_{k, \rm{el-n}}$  is defined as in Sec.~\ref{sec::hf_structure} for $k \in \{x, y, z\}$. It follows that the nuclear spin flip probability is,
\begin{equation}
    \mathbb{P}_n^{j \rightarrow i} = (1-\delta_{ij}) \frac{\sum_{\mu} \mathcal{P}^{A,j \rightarrow \mu, i}}{\sum_{i} \sum_{j} \sum_{\mu} \mathcal{P}^{A,j \rightarrow \mu, i}}.
\end{equation} In Fig.~\ref{fig:spin_flip}(b), we plot the simulated PLE intensity, $I(t)$, during the nuclear spin initialization protocol. This can be computed as,

\begin{equation}
    I(t) = \frac{1}{2} \gamma_e \sum_{j} P_{\rm{exc}}^j P_j.
\end{equation} We see that an exponential decay in $I(t)$ is observed, from which we compute the initialization rate. The simulated initialization rates plotted in the Fig.~4(b) of the main text, use the the nuclear spin flip probabilities shown in Fig.~\ref{fig:spin_flip}. The parameters used in the simulation are: $\Omega_1/\sqrt{p} = 191$~GHz/W\textsuperscript{1/2}, $\Omega_2/\sqrt{p} = 47$~GHz/W\textsuperscript{1/2}, $p_{\rm{sat}} = 70$~nW, $\tau_r = 5.9$~ns, $\Gamma_0 = 600$~MHz, $T_2^* = 200$~ns, $\omega_{\rm{hf}}/2\pi = 34$~MHz, $A_{g, \parallel} = A_{g, \perp} = 37$~MHz, $A_{u, \parallel} = 19$~MHz and $A_{u, \perp} = 10$~MHz.

\section{Nuclear Polarization Lifetime}

\begin{figure}[t]
\centering
\includegraphics{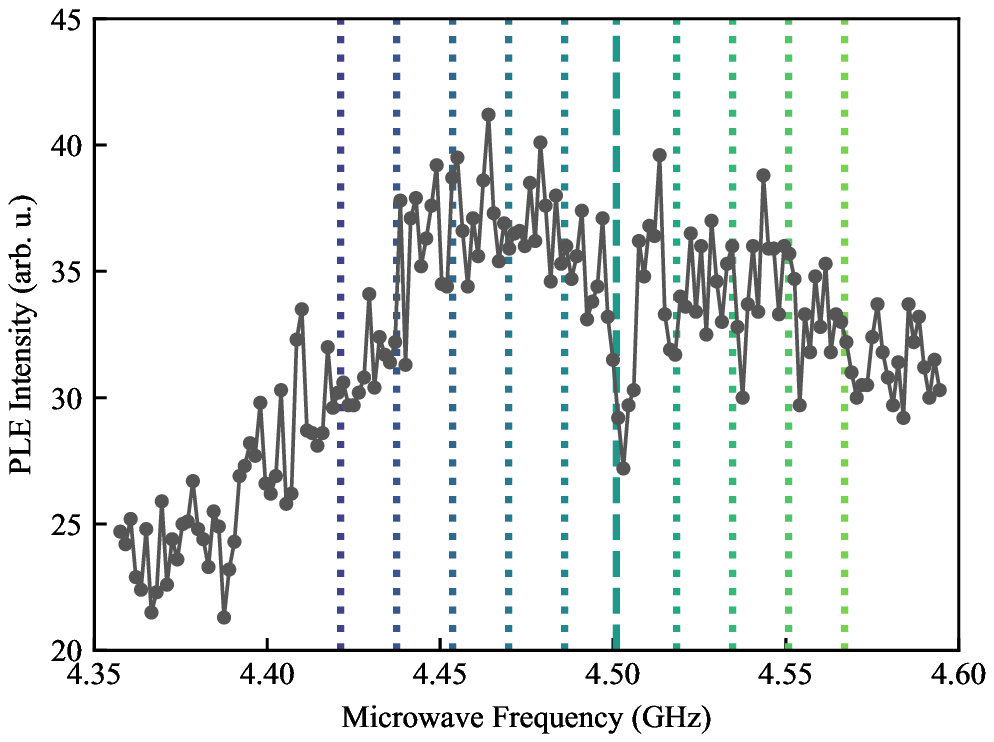}
\caption{\label{fig:nT1} PLE intensity measured during the first $50$~ns of a CPT\textsubscript{probe} pulse. The CPT\textsubscript{pump} frequency (indicated by the teal dashed line) is resonant to the CPT dip corresponding to the nuclear $\ket{1/2}$ state. A wait time of $100$~$\mu$s separates the CPT\textsubscript{pump} and CPT\textsubscript{probe} pulses. The CPT resonances that were not initialized by the CPT$_{\rm pump}$ pulse are indicated by the coloured dotted lines.}

\end{figure}

To rule out a short lived nuclear spin lifetime, we perform a nuclear pump-probe experiment with a $100$~$\mu$s wait time between the CPT\textsubscript{pump} and CPT\textsubscript{probe} pulses. The result - plotted in Fig. \ref{fig:nT1} - shows a single single CPT dip corresponding to the pumped nuclear state (in this case $\ket{1/2}$). Dips corresponding to the other resonances are not visible up to the measurement noise. From this we place a lower bound on the nuclear spin relaxation time,  $T_{1,n} > 100$~$\mu$s and conclude that the nuclear spin randomization occurs due to optical excitation rather than longitudinal relaxation.

\bibliography{suppreferences}